\documentclass[pdflatex,sn-nature,iicol]{sn-jnl}

\usepackage[utf8]{inputenc}
\usepackage[dvipsnames]{xcolor}

\usepackage{amsmath,amssymb,amsthm}
\usepackage{physics}
\usepackage{bm}
\usepackage{bbold}
\usepackage{mathrsfs}

\usepackage{graphicx}
\usepackage{dcolumn}
\usepackage{booktabs}
\usepackage{float}
\usepackage{subfigure}
\usepackage{afterpage}

\usepackage{tikz}
\usetikzlibrary{
  positioning,
  calc,
  fit,
  arrows.meta,
  decorations.pathreplacing,
  shapes.geometric
}

\usepackage{algorithm}
\usepackage{algorithmicx}
\usepackage{algpseudocode}

\usepackage{usebib}
\usepackage{soul}
\usepackage{subfiles}

\usepackage{bibunits}
\defaultbibliography{bibliography}

\hypersetup{
  colorlinks=true,
  linkcolor=NavyBlue,
  citecolor=NavyBlue,
  urlcolor=NavyBlue
}

\newcommand{\Ot}{\tilde{O}}

\providecommand{\melem}[3]{\langle #1 \vert #2 \vert #3 \rangle}

\theoremstyle{plain}

\theoremstyle{definition}

\theoremstyle{remark}

\tikzset{
  tens/.style = {draw, thick, rounded corners=1.5pt, fill=white,
                 minimum width=9mm, minimum height=7.5mm, inner sep=1.5pt,
                 font=\small},
  lbld/.style = {tens, fill=NavyBlue!12},
  mpo/.style  = {tens, fill=black!8},
  cap/.style  = {draw, thick, fill=Maroon!15, isosceles triangle,
                 isosceles triangle apex angle=60, minimum size=3.5mm,
                 inner sep=1pt, font=\scriptsize},
  leg/.style  = {thick},
  ghost/.style= {tens, densely dashed, fill=white},
  lab/.style  = {font=\scriptsize, inner sep=1.5pt},
}

\algrenewcommand\algorithmicrequire{\textbf{input:}}
\algrenewcommand\algorithmicensure{\textbf{output:}}

\tikzstyle{tensor}=[rectangle,draw=blue!50,fill=blue!20,thick]

\tikzstyle{H}=[rectangle,draw=red!50,fill=red!20,thick]

\tikzstyle{operator}=
[circle,draw=red!50,fill=red!20,thick]

\usepackage[normalem]{ulem}

\begin{document}
\begin{bibunit}[naturemag]

\title{Selective Control and Imaging of Higgs and Goldstone Modes with a Quantum Simulator}

\author[1,2]{\fnm{Connie} \sur{Miao}}
\equalcont{These authors contributed equally to this work.}

\author[4,2]{\fnm{Pranjal} \sur{Praneel}}
\equalcont{These authors contributed equally to this work.}

\author[3]{\fnm{Thomas G.} \sur{Kiely}}

\author[2]{\fnm{Eliott} \sur{Rosenberg}}

\author[1,2]{\fnm{Nicole S.} \sur{Ticea}}

\author[2]{\fnm{Pedram} \sur{Roushan}}

\author*[4]{\fnm{Erich} \sur{Mueller}}
\email{em256@cornell.edu}

\author*[2]{\fnm{Andre G.} \sur{Petukhov}}
\email{apetukhov@google.com}

\affil[1]{\orgdiv{Department of Physics}, \orgname{Stanford University}, \orgaddress{\city{Stanford}, \state{CA}, \country{USA}}}

\affil[2]{\orgname{Google Quantum AI}, \orgaddress{\city{Santa Barbara}, \state{CA}, \country{USA}}}

\affil[3]{\orgdiv{Kavli Institute for Theoretical Physics}, \orgname{University of California, Santa Barbara}, \orgaddress{\city{Santa Barbara}, \state{CA}, \country{USA}}}

\affil[4]{\orgdiv{Department of Physics}, \orgname{Cornell University}, \orgaddress{\city{Ithaca}, \state{NY}, \country{USA}}}

\date{\today} %

\abstract{
We realize the mesoscopic Bose-Hubbard model at unit-filling on a two-dimensional array of transmon qubits and selectively excite collective modes associated with amplitude or phase fluctuations of the superfluid order parameter. We identify these modes respectively as the Higgs and Goldstone modes using complementary local observables, extracted from measuring the number of particles on each site. Modulation of the coupling between sites excites the Higgs mode, which we detect via the density of doubly occupied sites. In contrast, modulation of the local chemical potential with an appropriately chosen spatial profile excites the lowest-frequency Goldstone mode, which we detect via changes in the local density. The two modes have different symmetries and do not couple to one another. By varying the system geometry, we characterize the impact of finite system size on the evolution of both modes.
In the thermodynamic limit, decay into a continuum of lower-energy excitations can obscure the Higgs response. In our mesoscopic system, however, the finite system size leads to spectral gaps and allows us to detect the Higgs mode as a coherent, long-lived excitation. Our work demonstrates how finite quantum simulators can isolate and probe physical phenomena that become difficult to resolve in the thermodynamic, large-scale limit.
}

\maketitle

One of the most profound principles in modern physics is that the laws governing a system can possess more symmetry than the state it ultimately adopts. For instance, ferromagnetic spins select a particular direction, even in the absence of an applied field. This phenomenon of spontaneous symmetry breaking underlies a vast range of physical effects, from magnetism and superconductivity to the Higgs mechanism in particle physics~\cite{ChaikinLubensky1995,PekkerVarma2015}. Spontaneous symmetry breaking is generally accompanied by collective modes, which emerge as fluctuations of an order parameter. For a superfluid, the order parameter is a complex function $\psi(r)$ whose spatial structure determines the wavefunction of the macroscopically occupied condensate mode. %
Gradients of the phase of $\psi(r)$ correspond to supercurrent, while its squared magnitude represents the local condensate density.
Despite the central role of the superfluid order parameter, its microscopic dynamics have remained largely hidden from view. Here, we resolve and visualize the dynamics of collective excitations associated, in the thermodynamic description, with Higgs and Goldstone modes of a lattice superfluid without requiring a direct measurement of the order parameter.

To probe a collective excitation, one requires both a means to excite it and a means to detect it. Both of these must couple strongly, but selectively. 
It is challenging to probe the Higgs amplitude mode in a superfluid or superconductor,
as this mode does not couple directly to traditional probes like electromagnetic fields.  
We devise probes that couple directly to this mode.
A second challenge is that the Higgs mode can decay into pairs of lower-energy Goldstone (phonon) modes, especially in two dimensions~\cite{Shimano2020, PodolskyAuerbachArovas2011, Pollet2012, Gazit2013}, which leads to broad and indistinct spectral features.
We overcome this difficulty by working with a mesoscopic system, where the multi-Goldstone continuum is replaced by a discrete set of modes. This discreteness suppresses decay and permits sharp, long-lived Higgs-like excitations instead of a broad, overdamped resonance. 

\begin{figure}[!t]
\centering
\includegraphics[width=\columnwidth]{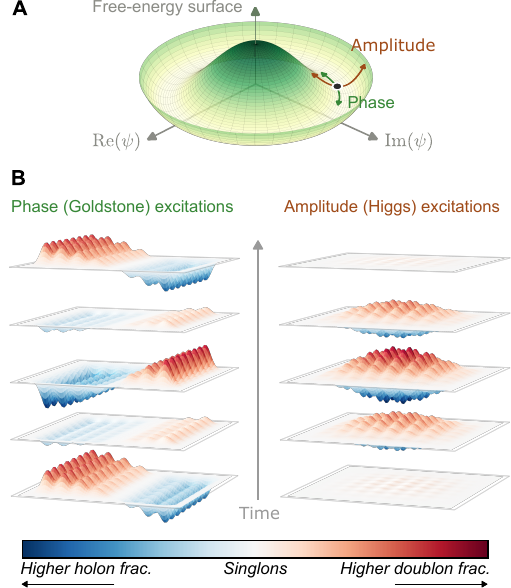}
\caption{\label{fig:fig1_mode_theory}
\textbf{Goldstone and Higgs oscillations.}  (\textbf{A}) Thermodynamic symmetry-broken representation of the free energy landscape in the superfluid phase. Spontaneous symmetry breaking yields a non-zero order parameter $\psi$ along the ring of minima. The exact finite, number-conserving state measured here retains the global $U(1)$ symmetry. (\textbf{B}) Goldstone modes manifest as density waves, whereas Higgs modes drive density variance fluctuations that oscillate between local doublon–holon pairs and singlon states, corresponding to local suppression/enhancement of the condensate fraction without changing the local density. Instantaneous probability of $\ket{2}$ (doublons), $\ket{0}$ (holons), and $\ket{1}$ (singlons) are shown in red, blue, and grey, respectively.}
\end{figure}

Experiments with ultracold atoms first detected the Higgs mode via Bragg spectroscopy~\cite{Bissbort2011} and the thermal response to lattice modulations~\cite{Endres2012}. These experiments found that spatial inhomogeneities in the trapping potential~\cite{Liu2015, Endres2012} and large drive strengths beyond linear response~\cite{Bissbort2011} broadened spectral features and complicated the modeling.
These challenges motivate our development of techniques for coherent control of the Higgs mode in the weak-driving regime on a spatially homogeneous system.
Here, we use a two-dimensional superconducting transmon lattice with qutrit readout to resolve spectra of the Higgs and Goldstone modes in the mesoscopic limit.
Individual-site readout reveals the microscopic local occupation dynamics of each mode.
Oscillations of the doublon--holon admixture provide a signature of the Higgs mode that does not require
direct measurement of the order parameter.
We demonstrate how mode interactions change with system size and show agreement with numerics at smaller system sizes.

The dynamics of microwave photons in coupled transmons can be mapped approximately onto the Bose–Hubbard model, governed by the Hamiltonian
\begin{multline}\label{eq:BoseHubb}
    H = -J\sum_{\expval{ i, j}}(a_i^\dag a_j + a_j^\dag a_i)
     + \frac{U}{2} \sum_i n_i (n_i-1)\\
     - \sum_i \mu_i n_i,
\end{multline}
where $a^\dagger_i = |1\rangle \langle 0| + \sqrt{2}\, |2\rangle \langle 1| + \dots$ creates a boson at site $i$ and $n_i\equiv a^\dagger_i a_i$.
Here, the coupling constant $J$ sets the hopping strength between nearest neighboring sites and is mediated by tunable coupling elements~\cite{Chen2014}. The on-site interaction strength $U$ is set by the transmon non-linearity, and the on-site potential $\mu_i$ is set by the qubit frequency detuning. 
We work in the unit-filled regime, where there is on average one particle per site. By only probing small $J/U$, where the on-site occupations can be limited to $n \in \{0,1,2\}$, the transmons may be treated as qutrits. 
Unlike an incommensurate hard-core model where $n \in \{0, 1\}$, this finite $U$ model retains doublon--holon fluctuations about the integer-filled state.
To calibrate our device for analog quantum simulation under this Hamiltonian with qutrit measurement, we follow the procedures outlined in \cite{Ticea2026, Andersen2025} (Appendix \ref{app:device_calibration}).
Recent experiments on transmon lattices have demonstrated many-body spectroscopy on the Bose-Hubbard model~\cite{robertsManybodyInterferometryQuantum2024, duTunnelingSpectroscopySuperconducting2025, Ticea2026} and spin-$1/2$ lattice models~\cite{andersenPrecisionQuantumSimulation2026, millarQuenchSpectroscopyMagnetic2026, leeBenchmarkingQuantumSimulation2026}. Our work focuses on the experimentally distinct regime of the clean, unit-filled Bose-Hubbard lattice. In this regime, we explicitly distinguish the Higgs and Goldstone modes within the superfluid phase, and we trace their behavior across the finite-size Mott--superfluid crossover.

The ground states of this model are governed by the competition between hopping $J$ and on-site repulsion $U$. For $J/U < g_T\equiv(J/U)_c$, the unit-filled ground state is a Mott insulator, and when $J/U\ll g_T$ it may be approximated as a product state with one boson per site.
The low-energy excitations in the Mott phase are particle and hole excitations; within the fixed-number sector they occur as pairs of doublons ($\ket{2}$) and holons ($\ket{0}$). For $J/U> g_T$, the ground state is a superfluid that spontaneously breaks the global U(1) symmetry. The superfluid-insulator transition at $g_T$ corresponds to a change in the free energy landscape of the complex order parameter, evolving from a symmetric paraboloid centered at $\psi(r)=0$ in the Mott phase to a ``Mexican hat" potential with a global minimum at $|\psi(r)|>0$ in the superfluid phase (Fig.~{\ref{fig:fig1_mode_theory}}A)~\cite{Fisher1989,Jaksch1998,Greiner2002}.
In the superfluid phase, the phase-like and amplitude-like oscillations of the order parameter on this potential surface form the two collective modes (Fig.~\ref{fig:fig1_mode_theory}B). Phase (Goldstone) modes propagate along the flat brim at constant amplitude, appearing spatially as a propagating sound wave in which the local density of photons $\langle n_i \rangle$ oscillates in time. In contrast, amplitude (Higgs) modes oscillate radially. The lattice ``breathes'' between configurations where all sites are synchronously more likely to host doublon-holon pairs and those where sites are predominantly singly occupied. This mode therefore couples to local density fluctuations rather than to the local density itself. Within mean field theory, oscillations of the amplitude of the order parameter are proportional to these oscillations of the doublon fraction~\cite{Altman2002}, see Appendix~\ref{app:mf_doublon}.
In a mesoscopic system much of the same physics persists, but the phase transition becomes a ``crossover regime'' and the quantum state typically involves a superposition of all symmetry-breaking configurations. For $J/U \gg g_T$ -- beyond the region that we probe -- we could no longer assume a maximum of two excitations per site, and this picture of amplitude and phase modes in terms of doublons and holons would break down~\cite{Altman2002}.

\begin{figure*}[t]
\centering
\includegraphics[width=\textwidth]{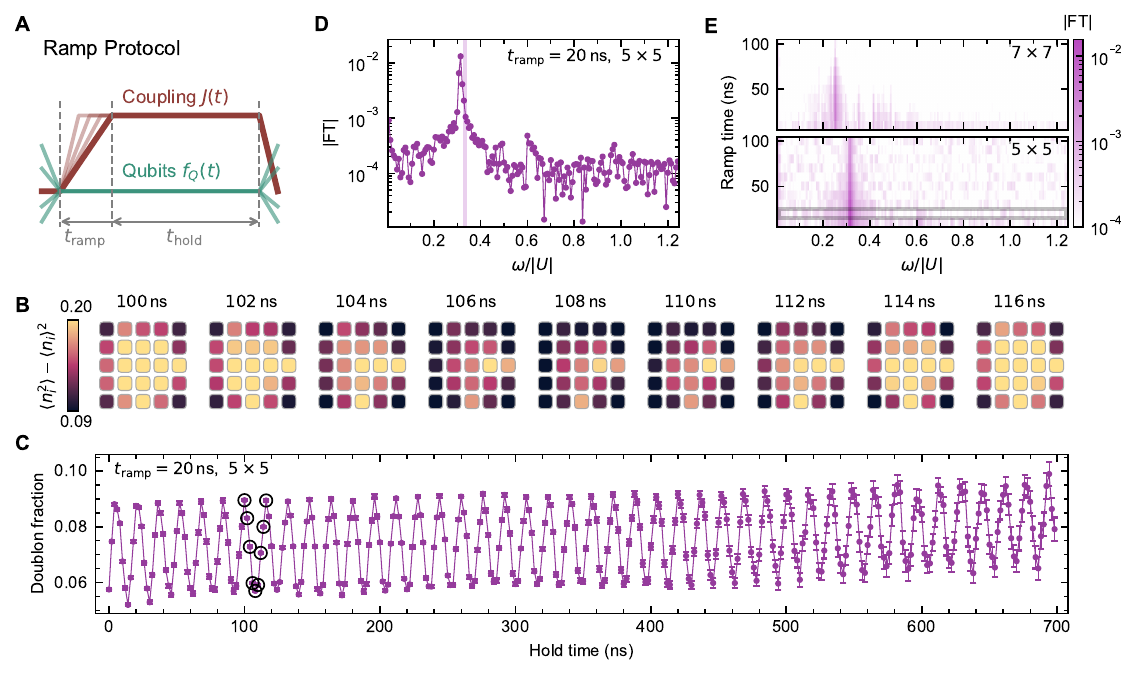}
\caption{\label{fig:fig2}
\textbf{Diabatic ramp spectroscopy of Higgs excitations.}
(\textbf{A}) Sketch of pulse protocol.
(\textbf{B}) Snapshots of density fluctuation for one period of the doublon fraction oscillation corresponding to circled points in (C).
(\textbf{C}) Measured oscillations of doublon fraction as a function of the hold time for $J/U=0.09$ and $\expval{U}/2\pi=-200.0 \pm 0.3$\,MHz. Plotted error bars represent 95\% confidence.
(\textbf{D}) Absolute value of the discrete Fourier transform of (C), showing that the maximum frequency component, corresponding to the lowest Higgs mode, is at $\omega/\abs{U}=0.314$. The theoretically calculated value is indicated by a vertical line.
(\textbf{E}) Fourier transform for varying ramp times, shown for both $7\times 7$ and $5 \times 5$ system sizes. The data corresponding to panels (C) and (D) is highlighted with a gray box.
}
\end{figure*}

\begin{figure*}[t]
\centering
\includegraphics[width=1.0\textwidth]{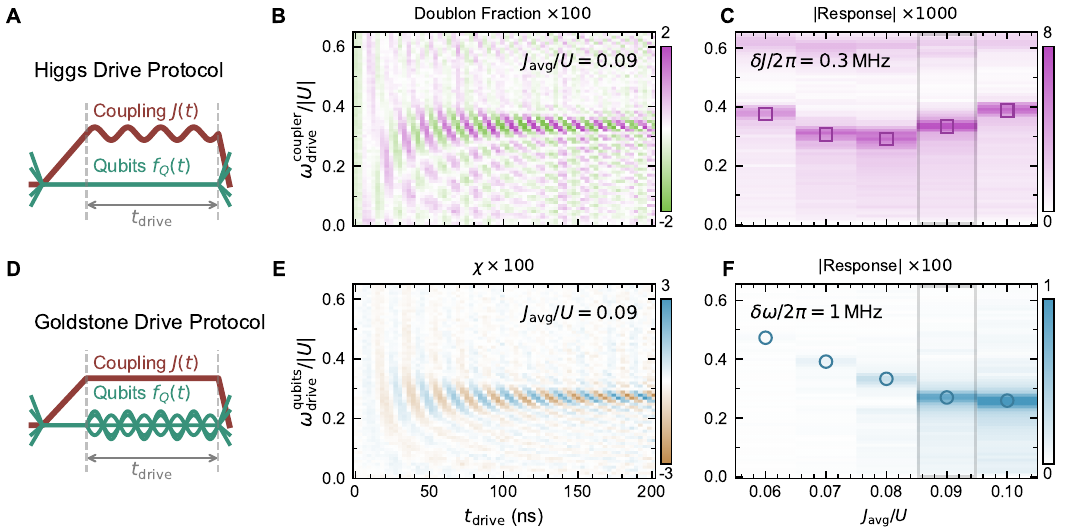}
\caption{\label{fig:fig3}
\textbf{Drive spectroscopy of Higgs and Goldstone excitations.}
We demonstrate coherent control to excite Higgs or Goldstone modes selectively on a $5\times 5$ system. 
(\textbf{A}) Pulse protocol to excite the Higgs mode.
(\textbf{B}) Measured relative doublon fraction as a function of time for different coupling drive frequencies; a best fit linear trend is subtracted from each row to highlight the oscillatory behavior. The corresponding Fourier transform is boxed in (C) in gray. 
(\textbf{C}) Response amplitude of the doublon fraction at each drive frequency as a function of the average coupling strength during the modulation. The Higgs response frequency, obtained via a fit to Lorentzian around the region of maximum response, is plotted as open squares. %
(\textbf{D}) Pulse protocol to excite the $k=1$ Goldstone mode. 
(\textbf{E}) Time traces vs. qubit modulation frequency of the density wave signal $\chi$; the corresponding Fourier transform is boxed in (F).
(\textbf{F}) Response amplitude of $\chi$ for modulation frequency $\omega_\text{drive}$ vs. the interaction coupling strength, with the $k=1$ Goldstone response frequency plotted as open circles.
}
\end{figure*}

\begin{figure*}[t]
\centering
\includegraphics[width=0.97\textwidth]{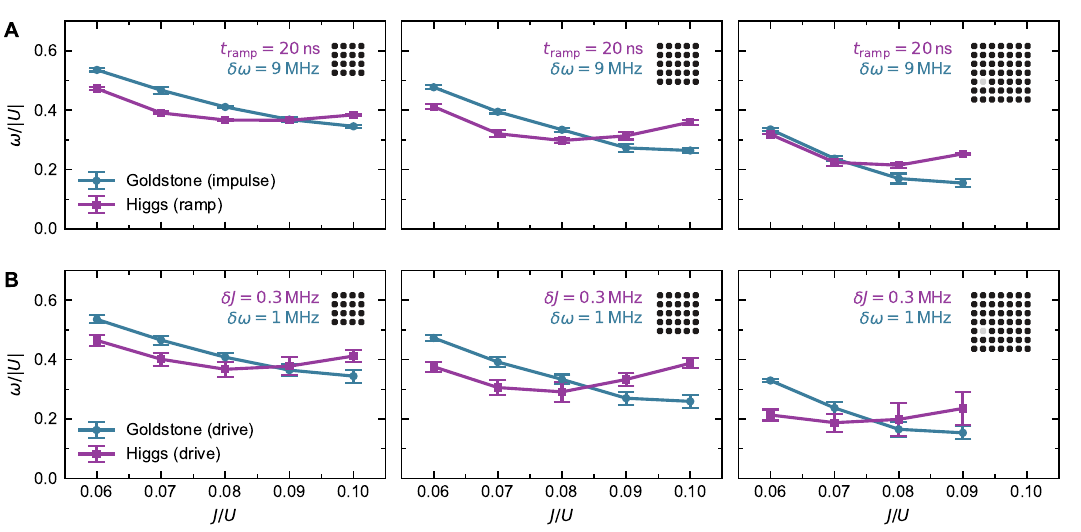}
\caption{\label{fig:fig4}
\textbf{System size dependency of Higgs and Goldstone modes.} (\textbf{A}) Vacuum Rabi spectroscopy: diabatic $J$-ramps to probe the Higgs spectrum (purple), and qubit frequency impulses to probe the $k=1$ Goldstone response (blue) for $4\times 4$, $5\times 5$, and $7\times 7$ (with one dropout qubit) lattice sizes, shown as insets. 
(\textbf{B}) Drive-based spectroscopy: modulation of $J$ to probe the Higgs response and qubit frequency modulation to probe the $k=1$ Goldstone response.
For all panels, we fit the peak response to a Lorentzian and plot the fitted frequency $\pm$ half the full width at half maximum.
}
\end{figure*}

To detect the Higgs mode, we set all $\mu_i=0$ and monitor oscillations in the doublon fraction $n_\text{doublon}/N$ -- the fraction of sites in $\ket{2}$ -- in response to a synchronous modulation of the hopping strength $J$ across all bonds.
The kinetic energy matrix element $\bra{h}K\ket{g}$, where $K=-J\sum_{\langle i,j \rangle} (a_i^\dagger a_j + \text{h.c.})$, couples the superfluid ground state $\ket{g}$ and the Higgs state $\ket{h}$ and allows us to detect the transition to the Higgs state~\cite{Endres2012, rubio-abadalFloquetPrethermalizationBoseHubbard2020}.
This coupling arises because modulating $J$ tunes the depth of the well formed by the ring of minima in the order parameter potential landscape, and the Higgs mode corresponds to the radial motion of the order parameter \cite{Fisher1989}. 
Since $\ket{g}$ and $\ket{h}$ are both eigenstates of the Hamiltonian $H=K+V$, we have $\bra{h}K\ket{g}=-\bra{h}V\ket{g}$, where $V=\frac{U}{2}\sum_i n_i(n_i-1)$ is the potential energy. Enforcing $\expval{n_i}=1$ with post-selection ensures the doublon fraction is proportional to $V$, so we can measure the doublon fraction to obtain information about the matrix element of $K$.
For a state prepared in a superposition $\ket{\psi(t)} = e^{-i E_g t/\hbar}\ket{g} + e^{-i E_h t/\hbar}\ket{h}$, the doublon fraction $\bra{\psi(t)} n_\text{doublon} \ket{\psi(t)}/N$ has a component that evolves with angular frequency $(E_h-E_g)/\hbar$, which reveals the  energy of the Higgs mode.
We therefore detect coherent excitation of the Higgs mode by observing the time-dependent oscillations of the doublon fraction.

\bmhead{Vacuum Rabi spectroscopy}
Recognizing the parallel between these oscillations and the oscillations between a coupled cavity and two-level system, we perform ``vacuum Rabi spectroscopy'' where we excite the Higgs mode and detect its frequency through a free-decay evolution.
In each experiment, we initialize the system at unit filling and bring the qubits onto resonance.
Since the Higgs mode is sensitive to a spatially uniform modulation of $J$, rapidly ramping the nearest neighbor couplings from zero (idle point) to a fixed value generates a Landau-Zener transition that prepares a superposition of the superfluid ground and Higgs mode states (Fig.~\ref{fig:fig2}A).
After a variable ramp time, we maintain a constant $J$ for a variable hold time, then perform a fast ramp down of both coupling and qubit frequencies to their idle points.
We then measure qutrit states to extract the doublon fraction, which oscillates as a function of the hold time.

Fig.~\ref{fig:fig2}B shows the spatial pattern of the number variance $\expval{n_i^2} - \expval{n_i}^2$ for nine hold times on a $5\times 5$ lattice. %
These patterns resemble drum-head modes, with some spatial structure that we attribute to differences in individual transmon properties such as readout fidelities, coherence times, non-linearities, or coupling strengths.
The spatial sum of these number variances is proportional to the doublon fraction, shown in Fig.~\ref{fig:fig2}C. %
At long times, post-selection reduces the number of surviving shots, leading to larger error bars (Appendix~\ref{app:doublon_decay} further discusses the effects of post-selection). 
The dependence of doublon fraction on hold time is dominated by a single frequency, as confirmed by the Fourier transform in Fig.~\ref{fig:fig2}D. The peak location agrees well with our theoretical modeling (vertical line).

We repeat this experiment for a range of ramp times and system sizes in Fig.~\ref{fig:fig2}E, using color to represent the amplitude spectrum of the doublon fraction response.
As the ramp time is shortened, the response amplitude grows, confirming that a more diabatic ramp more strongly populates the excited modes of the system.
The frequencies and widths of the peak response, however, are independent of the ramp rate, since the ramp rate affects only the state preparation weights and not their subsequent evolution.

By symmetry, our spatially uniform ramp of $J$ should not be able to excite the closest Goldstone mode.  %
Indeed, under this probe, we see negligible oscillations of the density $\expval{n_i}$, indicating that any such coupling is small (Appendix~\ref{app:selectivity}).
We can instead excite Goldstone modes by applying a short spatially inhomogeneous potential $\mu_i(t)$ and monitoring the free-decay oscillations of the density. 
We can choose the profile of $\mu_i(t)$ based on the momentum $q=k\pi/L$ for integer $k$ of the Goldstone mode that we want to excite, where the lattice size is $L\times L$.
In this work, we focus on the $k=1$ Goldstone mode, which is closest in energy to the Higgs.
The results of vacuum Rabi spectroscopy for this profile are shown in Appendix~\ref{app:goldstone_impulse_spec}.

\bmhead{Driven spectroscopy across phase transition}
Next, we demonstrate coherent driving between the superfluid ground state and each collective mode to map out the spectra across the insulator-superfluid phase transition on the $5\times 5$ lattice (Fig.~\ref{fig:fig3}). We again prepare the state by adiabatically ramping $J(t)$ from 0 to its final value, using the slowest ramp time shown in Fig.~\ref{fig:fig2}. We then sinusoidally modulate the couplings between all pairs synchronously 
as $J=J_\text{avg}+ \delta J \sin(\omega_{\text{drive}}t)$.
After time $t_\text{drive}$ we turn off the coupling and detune the qubits, then measure the doublon fraction $n_\text{doublon}(t)/N$.
As seen in Fig.~\ref{fig:fig3}B, the doublon fraction shows a resonant structure. We calculate the response function as the continuous Fourier transform evaluated at the drive frequency (Appendix~\ref{app:expt_data_analysis}).
In Fig.~\ref{fig:fig3}C we plot the absolute value of the response as a function of the coupling strength. Open squares show the resonant frequencies obtained by a Lorentzian fit to the maximum response in the data. The minimum near $J/U=0.08$ is a signature of proximity to the Mott-superfluid transition.

Similarly, we can coherently drive the Goldstone mode with momentum $q$ by modulating the qubit detunings instead of the couplings as $ \mu_{x, y}(t) = \delta\omega\,\sin(\omega_{\text{drive}} t)\cos(y q)$ (Fig.~\ref{fig:fig2}D).
Since the phase of the mode depends on the qubit position, we construct the signal as $\chi(t) = \sum_{x,y} N_{x,y}(t)\cos(y q) \times (2/L^2)$, shown in Fig.~\ref{fig:fig2}E.
The response is calculated using the Fourier transform of $\chi(t)$ taken at the drive frequency, shown in Fig.~\ref{fig:fig2}F.
We again see resonant behavior, with a single peak within the data-taking window. The fitted maximum response frequencies are shown in open circles. The mode frequency falls with increasing $J_\text{avg}$, and in the superfluid phase can be related to the speed of sound $c$ via $\omega=c\pi/L$. In the Mott phase the mode is gapped, and has a spectral weight which vanishes in the thermodynamic limit.

We note that the Higgs drive protocol excites a strong response in both the superfluid and insulating regime.
The Goldstone drive protocol, however, only produces a strong response in the superfluid regime.
This contrast is expected and is in fact a signature for both superfluidity and spontaneous symmetry breaking: in the Mott regime, the order parameter has zero magnitude and its phase is ill-defined. The amplitude mode, however, is still permitted.

Furthermore, the trends of each mode match with the thermodynamic limit behaviors.
We find that the Higgs mode reaches a minimum in the crossover regime, while the Goldstone mode approaches a constant frequency above this point.
In the thermodynamic limit, the Higgs mode frequency is non-zero in both the superfluid and insulating phases and reaches zero at the phase transition, while the Goldstone mode has zero energy in the entire superfluid regime. In finite size systems, the gap never closes to zero.

\bmhead{Finite size scaling}
There are no true phase transitions in finite systems -- in a finite-sized experiment with fixed particle number, the global U(1) symmetry cannot be broken. However, by repeating our experiments with different system sizes, we can track the approach toward the thermodynamic limit.
Such finite size scaling is common in numerical calculations~\cite{Fisher1989, geisslerFinitesizeScalingAnalysis2021}, but is less commonly explored experimentally.

In Fig.~\ref{fig:fig4}, we repeat our experiments to detect the peak response frequency for both collective modes as a function of coupling strength on $4\times 4$, $5\times 5$, and $7\times 7$ lattices. We removed one site from the $7\times 7$ grid due to poor swapping behavior, likely caused by a nearby mode at the interaction frequency.
In panel A, we measure the response using vacuum Rabi spectroscopy, exciting each mode and monitoring its free decay. In panel B, we instead monitor the resonant response to a sinusoidal drive. These two approaches agree quantitatively, except for the Higgs mode in the largest system size for small $J/U$. We attribute discrepancies to non-linear drive effects, which shift the frequencies of the drive spectroscopy peaks relative to the vacuum Rabi peaks. For further discussion, see Appendix \ref{app:heatmaps_comparison}.

For all three system sizes the Goldstone and Higgs mode frequencies cross as $J$ is varied. 
We do not resolve any avoided crossing within the experimental range and resolution, consistent with the modes belonging to different spatial symmetry sectors. While the Goldstone mode frequency decreases monotonically with $J$, the Higgs mode has a frequency minimum at a value of the coupling constant that can be taken to be an estimate of the phase boundary.

Based on the finite size scaling form expected from the Bose-Hubbard model symmetry class, the critical ratio $g^*$ where each collective mode reaches its minimum frequency should approach the thermodynamic limit $g_T$ as $g^{*}(L) - g_T \;\propto\; L^{-1/\nu}$, where $1/\nu \approx 1.49$ and $g_T\approx 0.06$~\cite{Fisher1989,Campostrini2006,Burovski2006, CapogrossoSansone2008}. Our data is consistent with this prediction, although confirming the expected critical exponent would require a finer coupling resolution and a larger range of system sizes.

\begin{figure}[!t]
\centering
\includegraphics[width=\columnwidth]{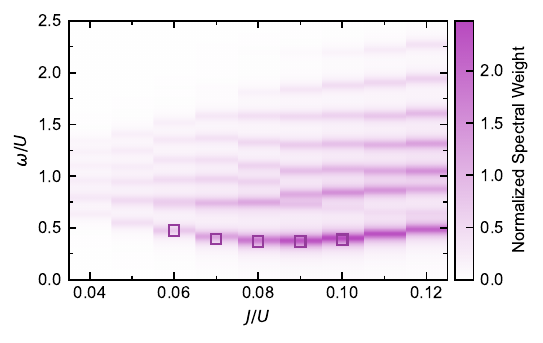}
\caption{\label{fig:fig_higgs_spectrum_theory}
\textbf{Simulated kinetic spectral function.} Unit normalized broadened spectral function (broadening $\eta = 0.05$)
calculated using tensor network techniques for a $4\times 4$ lattice.
Experimentally extracted frequencies from vacuum Rabi spectroscopy are shown as square markers.
}
\end{figure}

In the thermodynamic limit, the Higgs is broadened due to coupling to a continuum of states. In mesoscopic systems, this continuum separates into a finitely-spaced comb of discrete peaks, whose separation decreases as the system size increases. We numerically calculate the kinetic energy spectral function in Eq.~\eqref{eq:broadened_spectral} (Appendix~\ref{app:mesoscopic_to_thermodynamic}).
In Fig.~\ref{fig:fig_higgs_spectrum_theory}, we show a Lorentzian-broadened simulation of the Higgs-like states captured using numerics for a $4\times 4$ lattice. The multiple peaks are clearly visible. Repeating the calculation with smaller system sizes results in more broadly-spaced peaks. We show calculations using infinite matrix product states that illustrate their coalescence when broadened with the thermodynamic limit in Fig.~\ref{fig:Spectral_Function_Broadened} of the Appendix. The experiment probes the
first and most heavily weighted branch; others could be probed by increasing the drive frequency.

A fundamental open question in physics is whether distinct organizing principles govern behavior at intermediate, mesoscopic scales, bridging the gap between microscopic and thermodynamic extremes \cite{Laughlin2000}. Our results demonstrate that today’s quantum processors can directly address this challenge, serving as powerful tools to test whether -- and if so, how -- emergent phenomena adapt in finite-sized systems.

\backmatter

\bmhead{Acknowledgements}
We would like to thank Trond Andersen, Ilya Drozdov, Amir Karamlou, William Morong, Wojciech Mruczkiewicz, Richard Oliver, Alice Pagano, and Bicheng Ying for useful conversations and technical support.
C.M. acknowledges support from NSF GRFP Grant No. DGE-2146755.  E.J.M. and P.P. acknowledge support from NSF PHY-2409403. T.G.K. acknowledges support from the Gordon and Betty Moore Foundation under GBMF7392 and from the National Science Foundation under grant PHY-2309135 to the Kavli Institute for Theoretical Physics (KITP). This research has benefited from the Mitchel Postdoctoral Scholar Career Development Fund.

\putbib[bibliography] %
\end{bibunit}

\clearpage
\onecolumn %
\begin{appendices}
\begin{bibunit}[naturemag]
\section{Device and calibration procedures}\label{app:device_calibration}

\begin{figure*}[!htbp]
\centering
\includegraphics[width=\textwidth]{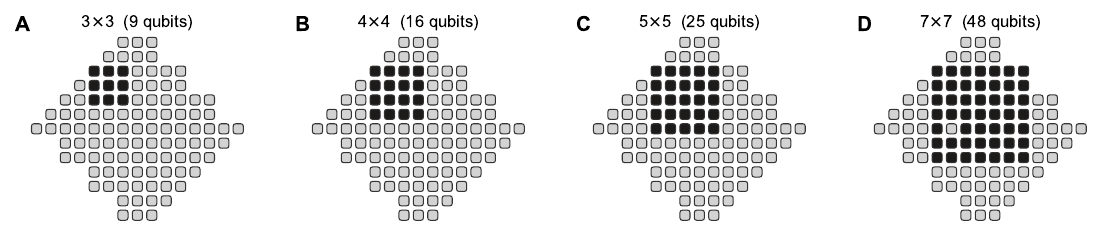}
\caption{\label{fig:system_sizes}
\textbf{Qubit lattices used in this work.} Light gray boxes show all data qubits in our device, while black boxes show the transmons used in our experiment. Couplers are not shown.
}
\end{figure*}

This work involves qubit lattices of several different sizes ($3\times 3$, $4\times 4$, $5 \times 5$, $7 \times 7$), with the data for the smallest, $3\times 3$, only shown in the appendices. All lattices are selected as a subset of the same device which hosts a larger grid of 105 qubits (Fig.~\ref{fig:system_sizes}). This device is similar to the one described in \cite{Ticea2026}, a superconducting architecture consisting of tunable couplers and frequency-tunable transmons arranged in a square lattice. To restrict the device to smaller lattice sizes, qubits and couplers are ``turned off'' by tuning them far out of range of the remaining qubits. The average anharmonicity of each grid of qubits at their idle point is $U/2\pi=-200.4 \pm 0.8$\,MHz, $-200.1 \pm 0.7$\,MHz, $-200.0 \pm 0.8$\,MHz, and $-199.8 \pm 1.0$\,MHz respectively for the four grids shown in Fig.~\ref{fig:system_sizes}, with error bars indicating the standard deviation. We treat $U/2\pi=-200$\,MHz as constant across all coupling strengths, although in practice the anharmonicity has a slight dependence on the coupling strength due to hybridization with the coupler~\cite{Ticea2026}. 
Although the device nonlinearity $U$ is negative, for a real Hamiltonian, complex conjugation maps evolution under $H$ to evolution under $-H$, preserving occupation probabilities for conjugate initial states. The correspondence with the repulsive model therefore requires consistent transformation of the preparation, potentials and drives, not a change in the sign of $U$ alone. See \cite{Ticea2026} for more details.

We follow the calibration procedure of \cite{Andersen2025}, described in detail below, which calibrates the qubit and coupler flux biases that generate the desired pair-wise coupling strengths $J_\text{target}$ between each $q_i$, $q_j$.
This procedure performs an ``analog'' calibration in the sense that it establishes the physical parameters that give rise to a fixed interaction Hamiltonian, instead of determining the parameters which govern digital gates on single qubits or pairs of qubits.
In this experiment, pairs of qubits are brought from their individual idle frequencies to a fixed target interaction frequency $f_\text{int}$, which is the same for all qubits. At the large coupling strengths (up to 20\,MHz) used in this work, qubits hybridize strongly with the coupler and experience much higher rates of decay compared to when idle or at lower coupling strengths. Furthermore, when the coupling is turned on, one poorly performing qubit can quickly act as a ``drain'' for photons from surrounding qubits. Thus, the interaction frequency must be carefully selected manually to ensure that all coupling calibrations can be completed successfully without being disturbed by nearby two level systems. 

We first define the variables that are relevant in the analog calibration. Each qubit and coupler can be biased with its own flux: $\Phi_{i,j}$ for qubits $i$, $j$, and $\Phi_c$ for the coupler. 
The two-qubit system has a Hamiltonian $H_{ij}= \omega_{q_i} a_i^\dag a_i + \omega_{q_j} a_j^\dag a_j + J (a_i^\dag a_j + a_j^\dag a_i)$.
The parameters $\omega_{q_i}$, $\omega_{q_j}$, and $J$ depend on all three flux set points in an inter-related way, and the purpose of the calibration is to empirically find the fluxes which lead to the desired coupling, $J$, and qubit frequencies, $\omega_{q_{i, j}}^{\rm on}=2\pi f_{\rm int}$.
Here, the superscript ``on'' refers to the fact that these are the qubit frequencies when the coupling is turned on.
For this two-qubit system the one-excitation sector has eigenstate energies $\omega_{\pm}= \frac{1}{2}\left(\omega_{q_i}+\omega_{q_j}\pm \sqrt{\delta^2+4J^2}\right)$, where $\delta=\omega_{q_j}-\omega_{q_i}$.  If one of the qubits is excited, the excitation swaps back and forth with rate $\Omega=\sqrt{\delta^2+4 J^2}$, which becomes $\Omega=2J$ if the fluxes are well chosen such that the detuning vanishes.

While \cite{Andersen2025} uses additional calibration steps to extract the bare qubit frequencies, the step that is most relevant for us is the iterative single photon ``swap spectroscopy'' step, which converges on the coupler and qubit biases to implement the desired coupling $J_\text{target}$. This experiment is equivalent to calibrating the two-qubit swap so that it is on-resonance with swap rate $\Omega = 2J$. We have 3 control parameters $\Phi_{i}$, $\Phi_{j}$ and $\Phi_c$, which need to be adjusted to achieve this set point. 
For each $\Phi_c$, we prepare one of the two qubits in $\ket{1}$, move both qubit flux bias points to achieve a variable qubit-to-qubit frequency detuning, then hold the coupler bias at the guess value $\Phi_c$ for a variable hold time to allow for swaps.
The qubit-to-qubit frequency detuning can be converted from the flux bias set points using a separate Ramsey experiment.
This 2D sweep results in the standard Rabi Chevron oscillation pattern, where we can fit the swap rate to $\Omega = \sqrt{\delta^2 + 4J_\text{achieved}^2}$ away from resonance. From the fit, we extract the achieved coupling $J_\text{achieved}$ for this coupler bias point, as well as the qubit flux bias points used to achieve $\delta=0$.

We then iterate on this experiment, progressively updating the coupler bias $\Phi_c$ and qubit flux bias sweep ranges based on how far the obtained $\Omega$ from the previous step is from the target, until $\Omega \approx 2J_\text{target}$ within a preset tolerance. The final qubit flux biases give us $\Phi_i^\text{on}$, $\Phi_j^\text{on}$ biases that are necessary to achieve $\delta = 0$ for the coupler bias value $\Phi_c$.

For each $J_\text{avg}$, we perform this iterative swap spectroscopy for each $J_\text{target} \in\{J_\text{avg},\,J_\text{avg} + \delta J,\, J_\text{avg} - \delta J\}$, with each calibration required to converge within an error of 0.02\,MHz for $\delta J=0.3$\,MHz. These target couplings are chosen so that we have a calibration point at each of the extrema in the coupling modulation used for the Higgs mode drive, between which we perform a linear interpolation in the coupling strength.
The iterative swap spectroscopy occasionally fails to converge on the target coupling. This can occur when there are nearby modes that distort the Rabi oscillation pattern and usually require a manual shift in the interaction frequency to fix.

We note that the pairwise-calibrated couplings and qubit frequencies are expected to shift slightly when other couplers are also turned on.
Indeed, with a few additional experiments it is possible to extract the underlying bare qubit frequencies, which could then be used to estimate the true frequencies and couplings achieved when all couplers are turned on, as was done in \cite{Andersen2025}. Based on the reasonably close match between our experimental spectra and the theoretical eigenenergies calculated with the pairwise-calibrated coupling values, we assume that the pairwise-calibrated coupling values are a reasonable approximation for the true effective coupling.

After the analog coupling calibration, each lattice is further calibrated following the procedure in \cite{Ticea2026} for the ability to do qutrit preparation and readout. The distribution of qubit idle frequencies must also be manually tweaked to ensure reasonable qutrit readout fidelities (additional TLS's can be encountered near the $f_{12}$ frequency, compared to only the $f_{01}$ frequency used for qubit readout) and to prevent transiting through regions with high photon loss rates on the way to the $f_\text{int}$.

In this work, the three smaller lattices use the same interaction and idle frequencies, while the $7\times 7$ uses a different set of frequencies for better performance on the larger grid. One qubit is removed from the $7\times 7$ grid due to anomalous behavior in the swap spectroscopy for all connected pairs, likely due to a nearby mode at the interaction frequency. We note that this defect only mildly perturbs the collective modes.

\section{Data analysis}\label{app:expt_data_analysis}
All data shown in this work is post-selected based on the total photon count. To compensate for the decay in surviving shots as a function of evolution time, we increase the number of shots for each time segment until shot noise decreases below the desired limit.

After binning each qutrit shot into one of the three possible Fock states, we calculate the doublon fraction per shot as the number of sites measured in $\ket{2}$ out of the total number of sites measured.
For the Higgs spectroscopy experiments, we fit the doublon fraction as a function of time to a linear trend line and subtract it to remove the low frequency component associated with the average doublon drift (Appendix~\ref{app:doublon_decay}); the time trace shown in Fig.~\ref{fig:fig2b} also has this linear trend subtracted. The Goldstone mode signal $\chi$ is calculated as
\begin{equation}\label{eq:chi}
    \chi = \sum_y (\chi_y - \bar{\chi}_y), \qquad
    \chi_y(t) = \sum_x \frac{2}{L} \frac{N_{x,y}(t)}{L}\cos(y\pi/L),
\end{equation}
where the lattice size is $L\times L$ and $N_{x,y}$ is the photon count at grid location $(x,y)$ and $\bar \chi_y$ is the time average.
In our model Hamiltonian we would expect $\bar\chi_y=0$, but it becomes non-zero in the presence of spatial inhomogeneities, such as due to different coherence times and readout fidelities, that break reflection symmetry. In our experiment this offset is nonzero.

To extract the Fourier response amplitudes in the vacuum Rabi spectroscopy shown in Figs.~\ref{fig:fig2}, \ref{fig:fig2b}, \ref{fig:impulse_heatmaps_system_sizes}, we calculate the frequency space response as a discrete Fourier transform. For the driven spectroscopy response amplitudes in Figs.~\ref{fig:fig3}, \ref{fig:drive_heatmaps_system_sizes}, we perform a continuous Fourier transform to demodulate the response at each drive frequency, where $\vec{d}$ is a vector of raw data (doublon fraction for the Higgs probe or $\chi$ for the Goldstone probe) of length $N_t$ corresponding to time points $\vec{t}$: $\text{response}(f) = \frac{1}{N_t} \vec{d} \cdot \exp(-i 2\pi f\vec{t})$.

To fit the four types of spectroscopy (Goldstone/Higgs with vacuum Rabi/drive spectroscopy), we use the same strategy of fitting a Lorentzian for a range of $\pm50$\,MHz about the frequency of maximum response, where the maximum response frequency is found for frequency values above $20\,$MHz to avoid finding a peak from low-frequency noise.

Large language models were used to aid in generating scripts for data analysis and processing.

\section{Doublon fraction drift and post selection}\label{app:doublon_decay}
\newcommand{\dP}[1]{\ensuremath{\dot{p}_{#1}}}
\newcommand{\p}[1]{\ensuremath{p_{#1}}}
\newcommand{\g}[1]{\ensuremath{\Gamma_{#1}}}

\begin{figure}[!ht]
\centering
\includegraphics[width=0.95\textwidth]{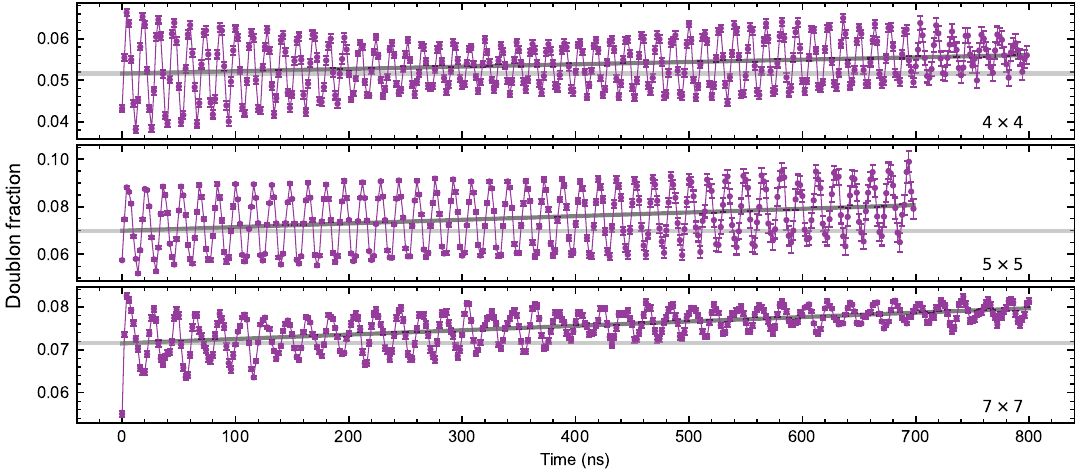}
\caption{\label{fig:detrend_demo}
\textbf{Rising doublon fraction for different system sizes.} Each dataset corresponds to a vacuum Rabi experiment with $J/U=0.09$ and $t_\text{ramp}=20$\,ns.
}
\end{figure}

\begin{figure}[!ht]
\centering
\includegraphics[width=\textwidth]{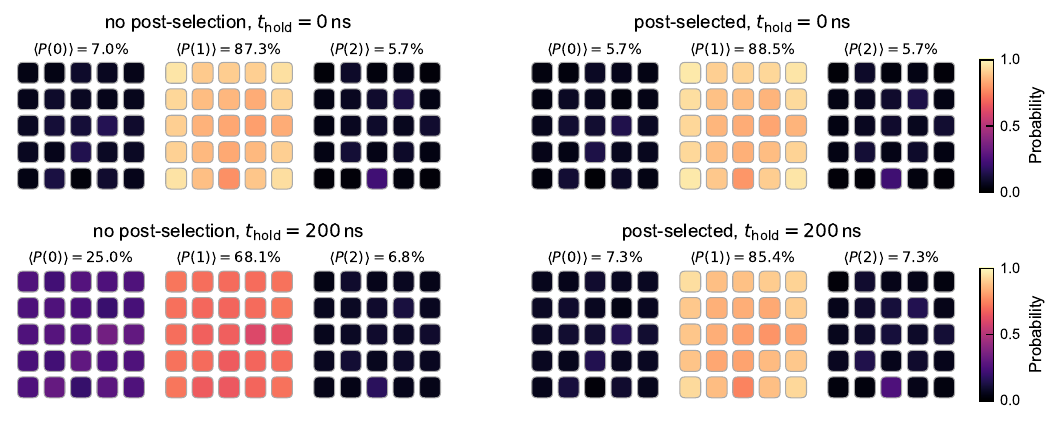}
\caption{\label{fig:post_selection_prob_grids_multi}
\textbf{Distribution of measured Fock states} with (left) and without (right) post selection on the $5\times 5$ lattice. We prepare an initial state using a slow coupling ramp (100\,ns) and measure the probability of each qubit being in $\ket{0}$, $\ket{1}$, or $\ket{2}$ at two different hold times (top and bottom rows). The useful comparison is the difference between columns rather than the absolute probabilities, since at different hold times we may be at a different point in the doublon fraction oscillation in Fig.~\ref{fig:detrend_demo}.
Here, the percentage of shots that survived post-selection out of 50,000 total shots is about 67\% at 0\,ns and 5\% at 200\,ns.
}
\end{figure}

In Fig.~\ref{fig:fig2}, the doublon fraction generally rises with evolution time.
Fig.~\ref{fig:detrend_demo} shows that the same trend appears across system sizes.
We show that two physically distinct effects can contribute to this effect.
First, $T_1$ loss removes shots from the fixed-number ensemble. Conditioning on the surviving shots will change the measured doublon fraction only if qubits have different decay rates, or if the singlon and doublon-holon configurations have different total decay rates. 
We show an example of the effects of post-selection at two different evolution times in Fig.~\ref{fig:post_selection_prob_grids_multi}.
Second, frequency noise acting on the dressed transmon--coupler eigenstates can drive transitions between singlon configurations and doublon--holon configurations without changing the total particle number. This mechanism produces incoherent doublons that survive postselection. 

The two-site models below isolate these two effects. The first explores when $T_1$ loss alone will bias the conditional (post-selected) distribution. The second shows how dephasing noise produces doublons within the number-conserving sector. More generally, any heating process will tend to increase the doublon fraction.

\subsection{Single qubit decay with post-selection}
Under certain conditions, single qubit decay can contribute to the increase in doublon fraction. 
Consider a simple picture with only two qubits, not necessarily neighboring, where we neglect any qubit-qubit coupling. We assume the qubit $a$ has decay rates $\g{1a}$ and $\g{2a}$ describing the $\ket{1}\to\ket{0}$ and $\ket{2}\to\ket{1}$ population decay channels respectively, and qubit $b$ has decay rates $\g{1b}$ and $\g{2b}$. We can solve for the probability $p_{ij}$ of the two qubits being in state $\ket{ij}$ as a function of time using a series of coupled differential equations:
\begin{gather}
    \dP{00} = \g{1b}\,\p{01} + \g{1a}\,\p{10}\\
    \dP{01} = \g{1a}\,\p{11} + \g{2b}\,\p{02} -\g{1b}\,\p{01}\\
    \dP{02} = -\g{2b}\,\p{02}\\
    \dP{10} = \g{1b}\,\p{11} + \g{2a}\,\p{20}-\g{1a}\,\p{10}\\
    \dP{11} = -(\g{1a} + \g{1b})\,\p{11}\\
    \dP{20} = -\g{2a}\,\p{20}
\end{gather}
\begin{gather}
    \p{00}(0)=\p{01}(0)=\p{10}(0)=0\label{eq:doublon_initial_cond0}\\
    \p{02}(0)=\p{20}(0)=(1-\p{11}(0))/2\equiv P_2\label{eq:doublon_initial_cond1}\\
    \sum_{i,j}\p{ij}=1. \label{eq:doublon_initial_cond2}
\end{gather}
To obtain the post selected probabilities, we must count only the states that are allowed by particle conservation ($\ket{11}$, $\ket{20}$, and $\ket{02}$; states with 3 excitations are assumed to never be reached). The probabilities that any one qubit has survived post selection and that it is measured in state $\ket{i}$ is then
\begin{equation}
    \p{0, \textrm{PS}} =\frac{1}{2}(\p{02}+\p{20}),\qquad 
    \p{1, \textrm{PS}} =\p{11},\qquad
    \p{2, \textrm{PS}} =\frac{1}{2}(\p{02}+\p{20}),\end{equation}
and the total probability of post selection survival is $\p{\textrm{PS}} = \p{0, \textrm{PS}} + \p{1, \textrm{PS}} + \p{2, \textrm{PS}}$.
The final post selected probabilities are $\p{i |\textrm{PS}} = \p{i, \textrm{PS}} / \p{\textrm{PS}}$.
Plugging in the solution to the coupled differential equations and defining $\g{i} \equiv (\g{ia}+\g{ib}) / 2$, $\delta\g{i} \equiv (\g{ia}-\g{ib})/2$, we find
\begin{gather}
    \p{0 |\textrm{PS}} = \p{2|\textrm{PS}} = \frac{2 - \epsilon}{2(2-\epsilon) + 4\xi e^{-(\g{\mathrm{eff}} + \delta\g{2})t}}, \qquad
    \p{1 |\textrm{PS}} = \frac{1}{1 + \frac{1}{\xi} e^{\g{\mathrm{eff}}t} \cosh(\delta\g{2}t)}\\
    \xi = \frac{1}{2P_2} - 1, \qquad
    \g{\mathrm{eff}} = 2\g{1} - \g{2}
    \qquad \epsilon = 1 - e^{-2\delta\g{2}t}
\end{gather}

First consider when $\delta\g{2}=0$ (note that there is no dependence on $\delta\g{1}$). In this case, if $P_2 < 1/2$ ($\xi > 0$), which is true in the experiment, we will have $\p{2,\textrm{PS}} > P_2$ for all time only if $\g{\mathrm{eff}}>0$, or in other words the decay rate $2\to1$ is less than twice as fast as the decay rate from $1\to 0$.
Intuitively, this condition makes sense because if the system is in $\ket{11}$, there are two channels for decay into states that will be thrown away ($\ket{10}$ and $\ket{01}$), whereas if the system is in a state with 2 excitations ($02$ or $20$), there is only one channel for decay into state that will be thrown away (also $10$ or $01$).
So, as long as the system reaches a discarded state from $\ket{11}$ \textit{faster} than it reaches the discarded state from $\ket{20}$, the post-selected system will appear to \textit{increase} in average doublon fraction.
Ideally, we would extract the decay rates $\g{ia}$, $\g{ib}$ experimentally.
However, these rates are difficult to measure accurately since when the coupling is turned on the decay rates change significantly due to hybridization between qubits and couplers. We instead estimate these rates from Fermi's golden rule, which says $\g{j\to k}\propto \rho(E_k)\abs{\bra{k}a\ket{j}}^2$ where $\rho(E_k)$ describes the density states around $\ket{k}$. Assuming the density of states is roughly the same at $\omega_1$ and $\omega_0$, we will have $\g{2} = 2\g{1}$. Thus, for transmons that experience uniform decay rates and a flat noise spectrum, the bias toward doublons is cancelled!

In the more experimentally-likely case, where the two qubits do not have the same $\g{1}$ decay rates, i.e. $\delta\g{2} \ne 0$, this condition on the relationship between $\g{2}$ and $\g{1}$ is relaxed. In this case, we would generate $\p{2,\textrm{PS}} > P_2$ for all time, even if both $\g{2a} = 2\g{1a}$ and $\g{2b} = 2\g{1b}$.

\subsection{Frequency noise within the two-excitation sector}
Frequency noise can change the doublon population without removing particles. To illustrate this mechanism, we consider two transmons, $a$ and $b$, connected through a tunable coupler $c$. We keep two excitations in total and at most one excitation in the coupler. The basis is $|200\rangle,|101\rangle,|002\rangle,|011\rangle,|110\rangle$, with occupations ordered as $(a,c,b)$.

We write Hamiltonians and energies in angular-frequency units, so that the physical Hamiltonian is $\hbar H$. The two qubits have frequency $\omega_{01}$ and anharmonicity $U/2\pi=-200\,$MHz; the coupler frequency is $\omega_{01}^c$. The direct qubit--qubit and qubit--coupler couplings are $g_{qq}$ and $g_{qc}$, respectively. Keeping the terms needed at the perturbative order specified below gives
\begin{equation}
\label{Eq:Hqqc}
H=\left(\begin{array}{ccccc}
2\omega_{01}+U&\sqrt{2}g_{qq}&0&0&\sqrt{2}g_{qc}\\
\sqrt{2}g_{qq}&2\omega_{01}&\sqrt{2}g_{qq}&g_{qc}&g_{qc}\\
0&\sqrt{2}g_{qq}&2\omega_{01}+U&\sqrt{2}g_{qc}&0\\
0&g_{qc}&\sqrt{2}g_{qc}&\omega_{01}+\omega_{01}^c&0\\
\sqrt{2}g_{qc}&g_{qc}&0&0&\omega_{01}+\omega_{01}^c
\end{array}\right).
\end{equation}
Here we define $\Delta=\omega_{01}^c-\omega_{01}$ and assume $|g_{qc}|\ll |\Delta|,|\Delta-U|$. We count $g_{qq}$ as being of the same order as $g_{qc}^2/\Delta$ and $g_{qc}^2/(\Delta-U)$. Direct qubit hopping between $|011\rangle$ and $|110\rangle$ then changes the effective Hamiltonian only beyond second order and is omitted in Eq.~\eqref{Eq:Hqqc}.

It is convenient to define
\begin{equation}
p=\frac{\sqrt2 g_{qc}}{\Delta-U},\qquad
q=\frac{g_{qc}}{\Delta},\qquad
C=pq=\frac{\sqrt2 g_{qc}^2}{\Delta(\Delta-U)}.
\label{eq:noise_dressing_parameters}
\end{equation}
A Schrieffer--Wolff transformation~\cite{BravyiSchriefferWolff2011} gives three dressed states that are normalized and mutually orthogonal through second order:
\begin{align}
|\psi_{200}\rangle&\simeq \left(1-\frac{p^2}{2}\right)|200\rangle-p|110\rangle-\frac C2|101\rangle+\dots,
\label{Eq:20}\\
|\psi_{101}\rangle&\simeq (1-q^2)|101\rangle-q\left(|110\rangle+|011\rangle\right)-\frac C2\left(|200\rangle+|002\rangle\right)+\dots,
\label{Eq:11}\\
|\psi_{002}\rangle&\simeq \left(1-\frac{p^2}{2}\right)|002\rangle-p|011\rangle-\frac C2|101\rangle+\dots.
\label{Eq:02}
\end{align}
The omitted terms are of higher order within the five-state model. These dressed configurations are not generally energy eigenstates. In this basis, after subtracting the common energy $2\omega_{01}+E_S$, the effective Hamiltonian is
\begin{equation}
H_{\mathrm{eff}}=\begin{pmatrix}
U_{\mathrm{eff}}&\sqrt2 g_{\mathrm{eff}}&0\\
\sqrt2 g_{\mathrm{eff}}&0&\sqrt2 g_{\mathrm{eff}}\\
0&\sqrt2 g_{\mathrm{eff}}&U_{\mathrm{eff}}
\end{pmatrix},
\label{eq:noise_effective_H}
\end{equation}
where
\begin{align}
E_S&=-\frac{2g_{qc}^2}{\Delta},\qquad
E_D=U-\frac{2g_{qc}^2}{\Delta-U},\\
U_{\mathrm{eff}}&=E_D-E_S
=U-\frac{2g_{qc}^2}{\Delta-U}+\frac{2g_{qc}^2}{\Delta},
\label{eq:noise_Ueff}\\
g_{\mathrm{eff}}&=g_{qq}-\frac{g_{qc}^2}{2}
\left(\frac1\Delta+\frac1{\Delta-U}\right).
\label{eq:noise_geff}
\end{align}
Here $E_D$ and $E_S$ are the diagonal doublon--holon and singlon energies measured from $2\omega_{01}$. The small shift $U_{\mathrm{eff}}-U$ does not change the qualitative behavior discussed below, provided $U_{\mathrm{eff}}<0$. It may also be neglected in the transition rates if the noise spectrum varies little over this shift.

We first neglect $T_1$ decay and consider frequency fluctuations of the qubits and the coupler:
\begin{equation}
\label{Eq:deltaHc}
\delta H(t)=\sum_{i=a,b,c}\delta\omega_i(t)n_i,
\qquad n_i=a_i^\dagger a_i.
\end{equation}
Using Eqs.~\eqref{Eq:20}--\eqref{Eq:02}, the second-order matrix elements are, for either $D=200$ or $D=002$,
\begin{align}
\label{Eq:2011}
\langle\psi_D|n_c|\psi_{101}\rangle&=C,\\
\langle\psi_D|n_a|\psi_{101}\rangle&=-\frac C2,\\
\langle\psi_D|n_b|\psi_{101}\rangle&=-\frac C2.
\end{align}
Their sum vanishes,
\begin{equation}
\langle\psi_D|(n_a+n_b+n_c)|\psi_{101}\rangle=0,
\end{equation}
as required for orthogonal states with the same total particle number. The noise matrix element is therefore
\begin{equation}
\langle\psi_D|\delta H|\psi_{101}\rangle
=C\left[\delta\omega_c-\frac{\delta\omega_a+\delta\omega_b}{2}\right].
\label{eq:noise_relative_detuning}
\end{equation}
This transition is driven by fluctuations of the coupler frequency relative to the mean qubit frequency. An identical shift of all three frequencies cannot induce transitions at fixed total particle number.

We define the unsymmetrized spectrum of angular-frequency noise as~\cite{clerkIntroductionQuantumNoise2010}
\begin{equation}
S_i(\omega)=\int_{-\infty}^{\infty}dt\,e^{i\omega t}
\langle\delta\omega_i(t)\delta\omega_i(0)\rangle.
\label{eq:noise_spectrum}
\end{equation}
For weak, independent noise sources, Fermi's golden rule gives transition rates between energy eigenstates~\cite{rasmussenSuperconductingCircuitCompanion2021,clerkIntroductionQuantumNoise2010}. To calculate these rates, we need to diagonalize $H_{\mathrm{eff}}V=V\,\mathrm{diag}(\epsilon_\nu)$ and transform the number operators to the new basis:
\begin{equation}
\Gamma_{\nu\rightarrow\mu}
=\sum_{i=a,b,c}\left|(V^\dagger n_i^{\mathrm{eff}}V)_{\mu\nu}\right|^2
S_i(\epsilon_\nu-\epsilon_\mu),
\label{eq:noise_eigenstate_rate}
\end{equation}
where $(n_i^{\mathrm{eff}})_{\alpha\beta}=\langle\psi_\alpha|n_i|\psi_\beta\rangle$ and $\epsilon_\nu$ is in angular-frequency units. This step includes the mixing of dressed configurations when $g_{\mathrm{eff}}\ne0$.

For a simple analytical example, set $g_{\mathrm{eff}}=0$ to the retained order. This removes coherent singlon--doublon mixing in Eq.~\eqref{eq:noise_effective_H}, although frequency noise can still drive transitions. %
The total rate from $\psi_{101}$ into the two doublon--holon configurations is $2\Gamma_{\downarrow}$, where
\begin{equation}
\Gamma_{\downarrow}=C^2\left[S_c(-U_{\mathrm{eff}})
+\frac14S_a(-U_{\mathrm{eff}})+\frac14S_b(-U_{\mathrm{eff}})\right].
\label{eq:noise_corrected_down_rate}
\end{equation}
For correlated noise, the bracket is replaced by the spectrum of $\delta\omega_c-(\delta\omega_a+\delta\omega_b)/2$, including cross correlations. Since $U_{\mathrm{eff}}<0$, the forward process releases energy to the noise bath and samples its positive-frequency spectrum. The reverse rate is obtained by replacing $S_i(-U_{\mathrm{eff}})$ with $S_i(U_{\mathrm{eff}})$, leaving $C^2$ unchanged. For a thermal bath,
\begin{equation}
S_i(-\omega)=e^{-\hbar\omega/k_BT}S_i(\omega),
\end{equation}
which relates the bath's ability to emit and absorb energy~\cite{clerkIntroductionQuantumNoise2010}.

To describe the dynamics with population rate equations, we also assume that coherence between $\psi_{200}$ and $\psi_{002}$ is suppressed by additional dephasing or a small, resolvable difference in their energies. Without this assumption, the perfectly symmetric system couples $\psi_{101}$ to $(\psi_{200}+\psi_{002})/\sqrt2$, with a matrix element larger by $\sqrt2$; the antisymmetric combination does not couple through this transition. The total forward rate is still $2\Gamma_{\downarrow}$, but the subsequent dynamics cannot in general be described by two independent population channels.

Under the population approximation, we take equal rates for the two channels,
$\Gamma_{101\rightarrow200}=\Gamma_{101\rightarrow002}=\Gamma_{\downarrow}$ and
$\Gamma_{200\rightarrow101}=\Gamma_{002\rightarrow101}=\Gamma_{\uparrow}$.
Any difference between the two transition frequencies is assumed small enough that their noise spectra and rates are approximately equal. We also neglect noise-induced transitions that populate the coupler, whose frequencies are of order $|\Delta|\gg |g_{qc}|,|U|$. Dropping the coupler index, the population equations are
\begin{align}
    \dP{11}&=-2\g{\downarrow}\,\p{11}+\g{\uparrow}\left(\p{20}+\p{02}\right)\label{Eq:RateEq1}\\
    \dP{20}&=\g{\downarrow}\,\p{11}-\g{\uparrow}\,\p{20}\label{Eq:RateEq2}\\
    \dP{02}&=\g{\downarrow}\,\p{11}-\g{\uparrow}\,\p{02}.\label{Eq:RateEq3}
\end{align}
We use the initial conditions in Eqs.~\eqref{eq:doublon_initial_cond0}--\eqref{eq:doublon_initial_cond2}. The stationary populations are
\begin{equation}
    \p{20}(\infty)=\p{02}(\infty)=\frac{\g{\downarrow}}{2\g{\downarrow}+\g{\uparrow}},\qquad
    \p{11}(\infty)=\frac{\g{\uparrow}}{2\g{\downarrow}+\g{\uparrow}}.
\end{equation}
To leading order in the small coupler admixture, the site-averaged doublon and holon fractions are $\p{2}(t)=\p{0}(t)=[\p{20}(t)+\p{02}(t)]/2$, and the singlon fraction is $\p{1}(t)=\p{11}(t)$. For a common thermal bath temperature, $\Gamma_{\uparrow}/\Gamma_{\downarrow}=e^{\hbar U_{\mathrm{eff}}/k_BT}$. At high temperature the two rates become equal, giving $\p{0}(\infty)=\p{1}(\infty)=\p{2}(\infty)=1/3$. As $T\to0$, the system instead approaches a doublon--holon configuration: $\p{2}(\infty)=\p{0}(\infty)=1/2$ and $\p{1}(\infty)=0$. Only one of the two sites is a doublon, even when the total probability of a doublon--holon configuration is one.

Nearly equal forward and reverse rates require $k_BT\gg\hbar|U_{\mathrm{eff}}|$ for a thermal bath. As a scale comparison, taking $U_{\mathrm{eff}}\simeq U$ gives $k_BT/h\simeq375\,$MHz at the refrigerator temperature of $18\,$mK, compared with $|U|/2\pi=200\,$MHz. The reverse-to-forward rate ratio is then about $e^{-200/375}=0.59$, so this is not yet the equal-rate limit. Effective qubit temperatures can be higher, for example $50$--$60\,$mK, but a temperature inferred from qubit populations need not describe the noise at the singlon--doublon transition frequency. Within this population model, an approximately symmetric noise spectrum drives an initially singlon-filled system toward equal probabilities of $\ket{11}$, $\ket{20}$, and $\ket{02}$.

Frequency noise therefore provides a possible explanation for the observed growth of the doublon population without particle loss. This mechanism differs from the postselection bias caused by $T_1$ decay. If the total loss rates from $\ket{11}$, $\ket{20}$, and $\ket{02}$ are equal, postselection leaves their relative populations unchanged. For identical qubits with $\Gamma_{2\to1}=2\Gamma_{1\to0}$, these rates satisfy $\Gamma_{11}=2\Gamma_{1\to0}=\Gamma_{20}=\Gamma_{02}$. Different qubit lifetimes, deviations from this level-scaling relation, or coupler loss can bias the measured drift. The two-site model gives a qualitative description of the mechanism, not a quantitative prediction of the observed slope.

\section{Drive amplitude sweep}

\begin{figure}[!htbp]
\centering
\includegraphics[width=0.4\textwidth]{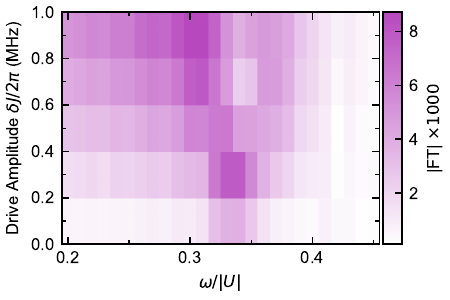}
\caption{\label{fig:drive_amp_sweep}
\textbf{Drive amplitude sweep for Higgs response} at $J/U=0.09$, system size $5\times 5$.
}
\end{figure}

To experimentally demonstrate that we are near the linear response regime in our Higgs drive spectroscopy protocol, we sweep the drive amplitude for drive frequencies around the Higgs response frequency for $J/U=0.09$ on the $5\times 5$ system size, shown in Fig.~\ref{fig:drive_amp_sweep}.
Here, we see that when the drive amplitude is increased three times from 0.1 to 0.3, the response increases (the maximum response increases by about a factor of two) without changing its frequency-dependent shape. On the other hand, when the response is increased another two times from 0.3 to 0.6, the response not only does not increase significantly but also seems to shift and broaden, which continues to be the case as the drive amplitude is further increased. This trend suggests that for the $5\times 5$ system, drive amplitudes at around $\delta J/2\pi \lesssim 0.3$\,MHz take us just outside the linear response regime, but is below the regime where obvious distortions begin to occur due to non-linear drive responses.

After selecting $\delta J/2\pi = 0.3$\,MHz to achieve the largest signal-to-noise ratio while still staying near the linear response regime of the $5\times 5$ system, we use the same $\delta J$ value for all system sizes in this paper.

\section{Goldstone vacuum Rabi spectroscopy}
\label{app:goldstone_impulse_spec}

\begin{figure}[H]
\centering
\includegraphics[width=\textwidth]{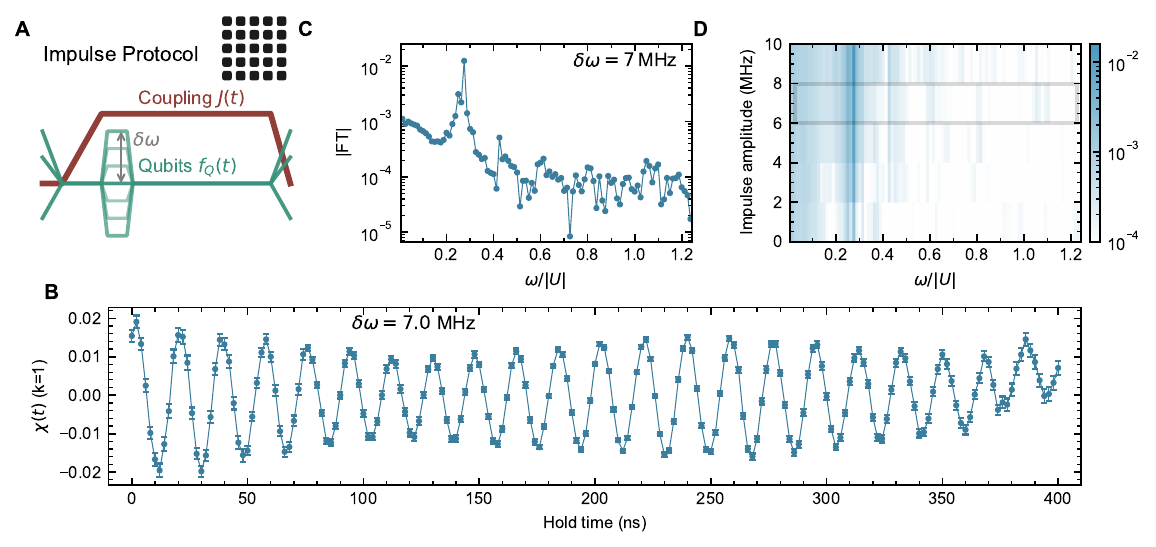}
\caption{\label{fig:fig2b}
\textbf{Impulse spectroscopy to probe $k=1$ Goldstone excitations at $J/U=0.09$.}
(\textbf{A}) Sketch of pulse protocol. Analogously to Fig.~\ref{fig:fig2}, the $k$-th Goldstone mode is excited by a fast ramp of qubit frequencies to $\delta\omega\,\cos(y\pi k / L)$ then back to the interaction frequencies. Varying the impulse amplitude generates different weights on the excited Goldstone modes. Here we show data for the $5\times 5$ system.
(\textbf{B}) Time trace of the response $\chi(t)$ with impulse amplitude $\delta\omega=7$\,MHz, calculated by demodulating each row of qubits' response according to its weight in the Goldstone mode's spatial profile (Eq. \eqref{eq:chi}). Error bars show 96\% confidence. (\textbf{C}) Absolute value of the Fourier transform of the time trace in (B). The maximum response is at about 55\,MHz. (\textbf{D}) $\chi$ response at varying impulse amplitudes. The trace in (C) is highlighted with a grey rectangle.
}
\end{figure}

Analogous to the ramp spectroscopy of Higgs modes, we can instead prepare a superposition of the superfluid ground state and Goldstone modes and measure the vacuum Rabi oscillations via an impulse procedure.
The Goldstone mode couples directly to the on-site density and is easy to address via individual qubit frequencies.
Following \cite{Ticea2026}, we can in general excite the Goldstone mode by bringing all qubits onto resonance at frequency $\omega_\text{int}$ and modulate the frequency detuning $\mu$ of the qubit at grid location $(x, y)$ with amplitude $\delta\omega$ and envelope $F(t)$:
\begin{equation}\label{eq:Goldstone_drive}
    \mu_{x, y}(t) = \delta\omega\,F(t)\cos(y \pi k / L)
\end{equation}
For the vacuum Rabi spectroscopy, $F(t)$ has a ramp up of 1\,ns, a hold of 5\,ns, and a ramp down for 1\,ns (Fig.~\ref{fig:fig2b}A). Amplitudes are scaled per qubit according to which row each qubit is located to implement a momentum kick along the $y$ direction. Once the qubit frequencies have returned to the interaction frequency, we evolve the prepared initial superposition for a variable time and measure the density response and extract $\chi$ according to Eq.~\eqref{eq:chi}.
Fig.~\ref{fig:fig2b}B shows the time trace corresponding to a drive amplitude of $\delta\omega=7.0$\,MHz. A single dominant frequency is apparent. The absolute value of the Fourier transform, shown in panel C, has one large peak, which we interpret as the natural oscillation frequency of the lowest energy phonon mode.
Some weaker peaks are seen, which correspond to either higher modes or multi-phonon excitations. Panel D explores the dependence of the response on the amplitude of the impulse $\delta\omega$. For impulses in this range the observed oscillation increases in strength with the impulse amplitude, indicating that the impulse amplitude indeed controls the amplitude of excitation into the Goldstone mode. The frequency remains the same at increasing amplitudes, consistent with our understanding of these oscillations as vacuum Rabi oscillations of state preparations of varying amplitude in the Goldstone mode.

\section{Selectivity of drives}\label{app:selectivity}

Spatially varying potentials can excite density waves~\cite{DisorderMediatedSuperfluidity,Ticea2026}. In our experiment, the two drive profiles address different spatial symmetries: uniform hopping modulation is reflection-even, whereas the lowest standing-wave potential profile is reflection-odd in the ideal array. This distinction allows selective excitation; it is not a general prohibition against exciting an amplitude-like mode with a potential. Below we test the selectivity by measuring both readouts after each preparation procedure. The theoretical connection between hopping modulation and doublon readout is given in Appendix~\ref{app:hopping_doublon_readout}.

\begin{figure}[b]
    \centering
    \includegraphics[width=0.95\textwidth]{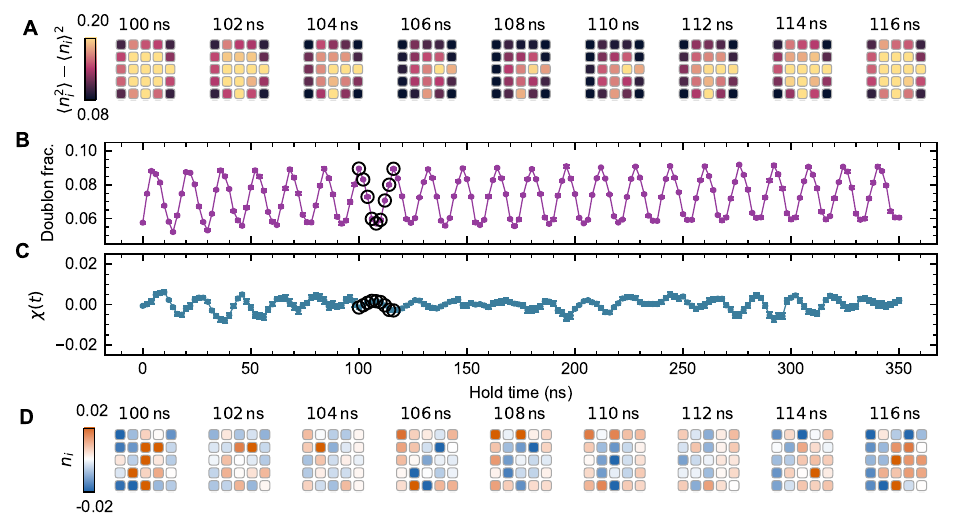}
    \caption{
    \textbf{Selectivity of diabatic coupler ramp procedure for Goldstone and Higgs modes}. \textbf{(A, B)} Replica of time trace and population variance snapshots as in Fig.~\ref{fig:fig2}. \textbf{(C)} Time trace of $\chi$ for $k_y=1$ calculated for the same dataset. \textbf{(D)} Snapshots of per-site population density $n_i$ after subtracting off each qubit's time-averaged density.
    }\label{fig:g_hold_selectivity}
\end{figure}

\begin{figure}[t]
    \centering
    \includegraphics[width=0.95\textwidth]{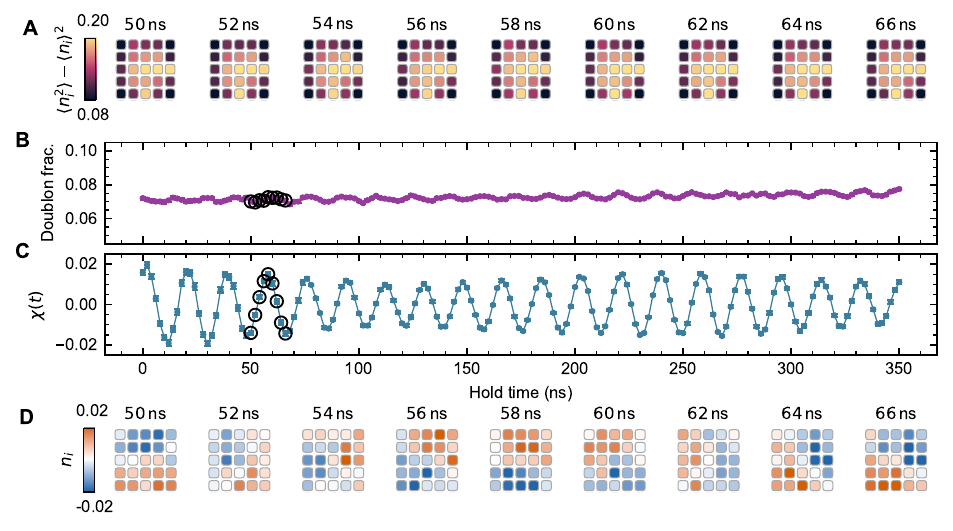}
    \caption{
    \textbf{Selectivity of frequency impulse procedure for Goldstone and Higgs modes}. Same as Fig.~\ref{fig:g_hold_selectivity} except using the $k_y=1$ profile frequency impulse for state preparation, with $\delta\omega=7$\,MHz. \textbf{(C)} is replicated from Fig.~\ref{fig:fig2b}.
    }\label{fig:w_impulse_selectivity}
\end{figure}

We observe that the two drive procedures, one modulating couplers synchronously and the other modulating qubit frequencies according to a $k=1$ profile, selectively excite either the amplitude or the phase-like modes and are protected via symmetry from exciting each other. Here we show the underlying data which demonstrates the selectivity of each drive. When applying the diabatic coupler ramp procedure (Fig.~\ref{fig:g_hold_selectivity}), we find a negligible response corresponding to a $k=1$ $\chi$ value. Similarly, when apply the frequency impulse procedure (Fig.~\ref{fig:w_impulse_selectivity}), we see a negligible signal in the Higgs response, while the density performs a standing-wave like oscillation between the top half and bottom half of the grid. The colorbar and y-axis scales are matched between Figs.~\ref{fig:g_hold_selectivity} and \ref{fig:w_impulse_selectivity} for easier comparison.

We note that the doublon fraction in Fig.~\ref{fig:w_impulse_selectivity}B, where we expect there is little to no weight on the amplitude modes, rests about halfway in between the two extremes of doublon fraction observed when the Higgs mode is excited (i.e. in Fig.~\ref{fig:g_hold_selectivity}B). This observation emphasizes that in this experiment the Higgs mode is signaled by an oscillating doublon fraction, not an elevated doublon fraction. Similarly the Goldstone mode is signaled by an oscillating density pattern, not a static distortion.

\section{Full spectroscopy data with theoretical comparison}\label{app:heatmaps_comparison}

In the main text we presented the data for vacuum Rabi spectroscopy for the Higgs mode on a 5x5 grid for a single value of $J/U$ (Fig.~\ref{fig:fig2}). We also showed the $J/U$ dependence of the fitted peak response frequencies from drive spectroscopy on both the Higgs and Goldstone modes on a 5x5 grid in Fig.~\ref{fig:fig3}.
Here we show the full spectroscopic data for a wider range of system sizes and parameters.

\begin{figure}[p]
    \centering
    \includegraphics[width=\textwidth]{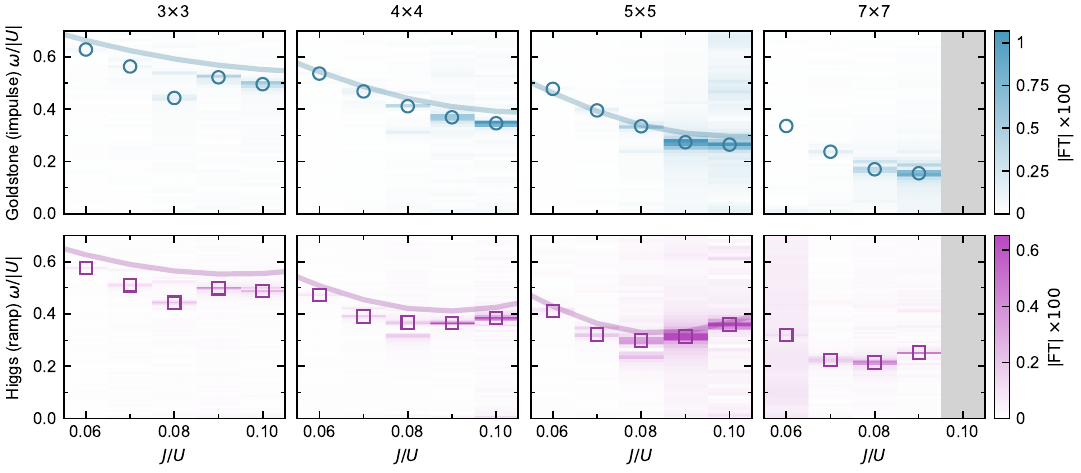}
    \caption{\label{fig:impulse_heatmaps_system_sizes}
    \textbf{Vacuum Rabi spectroscopy for Goldstone and Higgs modes.}
    Frequency response measured by monitoring the free oscillations after either (top) adding an impulsive potential whose spatial structure matches the lowest Goldstone mode or (bottom) ramping the coupling constant to excite the amplitude mode. Darker colors represent larger response. The fitted peak location of the largest response is shown in open circle or square markers.  Smooth curves show the bare eigenenergies extracted from our theoretical modeling.
    }
\end{figure}

Fig.~\ref{fig:impulse_heatmaps_system_sizes} shows the vacuum Rabi spectroscopy for each mode in the $3\times 3$, $4\times 4$, $5\times 5$, and $7\times 7$ lattice (the $7\times 7$ lattice has one qubit deactivated as described in Appendix \ref{app:device_calibration}).
Fig.~\ref{fig:drive_heatmaps_system_sizes} shows the driven spectroscopy for each mode in the $4\times4$, $5\times5$ and $7\times7$ lattices.
Each heatmap includes markers (open circles and squares) corresponding to the fitted frequencies of the highest amplitude peak. 
We show theoretical lines corresponding to the expected eigenenergies on the vacuum Rabi spectroscopy heatmaps for the three smaller system sizes. The theoretical energies are generally consistent with our experimental results, and we attribute the discrepancies to our uncertainty in model parameters.
For the smallest ($3\times 3$) system, there is a notable splitting in the maximum response frequency at $J/U=0.08$, particularly in the vacuum Rabi spectroscopy, with the maximum response dipping significantly below the theoretical energy.
We believe this splitting is related to non-linear effects. In particular, although the vacuum Rabi spectroscopy cannot display AC-Stark-type shifts since no drive is applied during the integration time over the response, the diabatic ramps used are really broadband frequency modulations that could prepare a superposition that includes population in higher states.
We hypothesize that this is the reason for the splitting to be particularly prominent in the vacuum Rabi spectroscopy.
As discussed in the main text, each system size demonstrates qualitatively similar features, and the two techniques for spectroscopy lead to similar excitation frequencies.

\clearpage
\begin{figure}[p]
    \centering
    \includegraphics[width=0.8\textwidth]{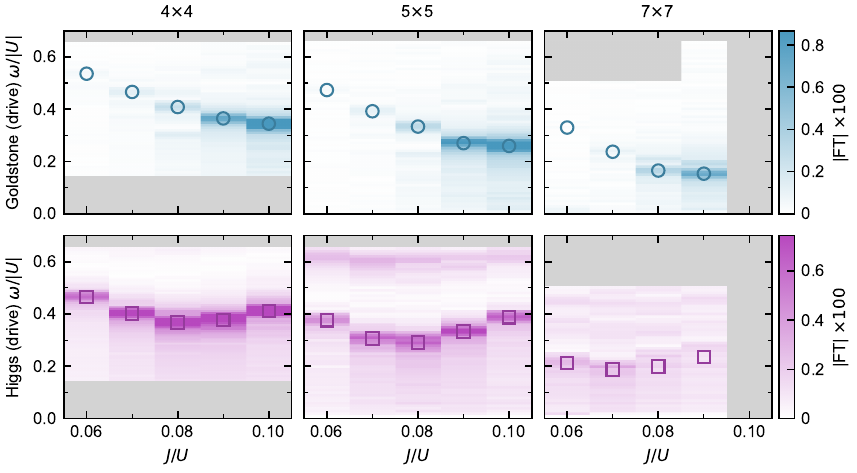}
    \caption{\textbf{Driven spectroscopy for Goldstone and Higgs modes.}}\label{fig:drive_heatmaps_system_sizes}
\end{figure}
\clearpage

\section{Collective modes and the finite-system crossover}
\label{app:mf_interpretation}

The superfluid description of the Bose--Hubbard model contains a phonon
branch associated mainly with phase fluctuations of the order
parameter~\cite{PhysRevB.75.085106,PodolskyAuerbachArovas2011}. Near the
superfluid side of the commensurate transition, the finite-interaction
model also supports amplitude-like oscillations involving coherent
changes in the doublon--holon admixture~\cite{Altman2002,PodolskySachdev2012,PekkerVarma2015}.
This distinction concerns the commensurate Bose--Hubbard model, rather
than an incommensurate hard-core model.

We first explain the mean-field connection between doublon fraction and
order-parameter magnitude, then discuss the superfluid-like crossover in
a finite, fixed-number system. Finally, we derive the relation between
hopping modulation and the measured doublon oscillation. The identities
used in this last step do not require mean field or broken U(1) symmetry.

\subsection{Doublon fraction and order-parameter magnitude}
\label{app:mf_doublon}
A simple three-level Gutzwiller state makes the connection to the
occupation readout explicit, in the spirit of Ref.~\cite{Altman2002},
while retaining the physical boson matrix elements $1$ and $\sqrt2$.
For a uniform local mean density of one, $|f_0|^2=|f_2|^2=r$ and
$|f_1|^2=1-2r$, where $r$ is the doublon fraction and $f_i$ is the amplitude associated with being in state $i$. A normalized local state, with the overall phase fixed, is
\begin{equation}
|f\rangle=\sqrt r\,e^{i(\chi-\varphi)}|0\rangle+
\sqrt{1-2r}\,|1\rangle+
\sqrt r\,e^{i(\chi+\varphi)}|2\rangle,
\qquad 0\leq r\leq\tfrac12.
\label{eq:commensurate_qutrit}
\end{equation}
Here $\varphi$ is the particle-phase angle and $\chi$ is a relative phase
of the doublon--holon admixture. Direct evaluation of the order parameter gives
\begin{align}
\psi=\langle a\rangle
&=e^{i\varphi}\sqrt{r(1-2r)}
       \left(e^{-i\chi}+\sqrt2\,e^{i\chi}\right),\nonumber\\
|\psi|^2&=r(1-2r)\left[3+2\sqrt2\cos(2\chi)\right].
\label{eq:mf_doublon_general}
\end{align}
The phase-aligned mean-field ground state has $\chi=0$.
Linearizing about $(r_0,\chi_0=0)$ in the ordered state yields
\begin{equation}
|\psi_0|^2=(3+2\sqrt2)r_0(1-2r_0),\qquad
\delta|\psi|=\frac{(3+2\sqrt2)(1-4r_0)}{2|\psi_0|}\,\delta r.
\label{eq:mf_doublon_linear}
\end{equation}
The relative-phase contribution to $|\psi|^2$ begins at second order at
this stationary point. Thus, for small coherent amplitude oscillations
and $0<r_0<1/4$, the doublon fraction and order-parameter magnitude
oscillate at the same frequency and in phase. %

Equation~\eqref{eq:mf_doublon_linear} is a mean-field relation, not a
model-independent inference from $r(t)$. An incoherent mixture can have
the same occupations but no $\langle a\rangle$. Moreover, the exact
fixed-number state has $\langle a_i\rangle=0$ even when its gauge-invariant
correlations are strong. For a finite inhomogeneous array, global unit
filling does not impose $\langle n_i\rangle=1$ at every site; the general
local expressions $\psi_i=f_{i0}^*f_{i1}+\sqrt2 f_{i1}^*f_{i2}$ and
$r_i=|f_{i2}|^2$ must then be used. A standing sound mode can also modulate
$r_i$ locally. The mode assignment therefore relies on the drive
symmetry, spatial response and frequency together, not on doublon
oscillations alone.

\subsection{Signatures of superfluid-like behavior in a mesoscopic system}
\label{app:finite_superfluid}
In an exact finite state with fixed total particle number,
$\langle a_i\rangle=0$ follows from global U(1)  symmetry. This does not
require the number-conserving correlations
$\rho^{(1)}_{ij}=\langle a_i^\dagger a_j\rangle$ to vanish, nor does it
exclude collective density and current motion. There is no unique sharp
Mott--superfluid boundary for these small arrays. The labels ``Mott-like''
and ``superfluid-like'' describe a crossover and its connection to bulk
behavior, rather than distinct symmetry sectors of the finite ground state.

A signature used here is a coherent, extended standing density
wave with the spatial structure of the lowest sound mode. In the
long-wavelength superfluid description its frequency obeys
$\omega_1\simeq c q_{\min}$, with $q_{\min}$ of order $\pi/L$ (lattice
spacing set to one); it is therefore nonzero at finite $L$ even when the
corresponding bulk branch is gapless. The exact allowed profile and
frequency depend on boundaries, inhomogeneity and proximity to the
crossover. The existence of a finite excitation gap is consequently
not, by itself, a criterion for a Mott-like state.

A single density resonance is not a sufficient universal test of
superfluidity: gapped insulators have density excitations, and sound can
also occur in normal fluids. The interpretation is strengthened by the
standing-wave profile, its evolution with size and coupling, and its
connection to the low-energy sound branch. A separate stiffness or
current-response measurement would provide a more direct transport test;
for example, on a ring or periodic lattice the ground-state energy
curvature under a boundary twist defines a phase stiffness.\footnote{
For a number-conserving finite-lattice comparison of stiffness with other
observables, see R.~Roth and K.~Burnett, Phys. Rev. A \textbf{67},
031602(R) (2003),
\href{https://doi.org/10.1103/PhysRevA.67.031602}{doi:10.1103/PhysRevA.67.031602}.}
A uniform static bond phase on a simply connected open array can instead
be removed by a gauge transformation, so the periodic-boundary twist
test cannot be transferred to this geometry without specifying boundary
constraints or the appropriate dynamical response. We do not report a
stiffness measurement here. The standing sound wave is a
signature of the superfluid-like crossover, not proof of symmetry breaking.

\subsection{Hopping modulation and doublon readout}
\label{app:hopping_doublon_readout}
\label{app:higgs_coupling_theory}
\label{app:number_fluctuations}
\label{app:coupling_higgs_probe}
We now relate the drive to the measured occupation signal without using
mean field. Let
\begin{equation}
P_{2,i}=|2\rangle_i\langle2|,\qquad
\hat r=\frac{1}{N_{\rm sites}}\sum_iP_{2,i},\qquad
r=\langle\hat r\rangle.
\end{equation}
For occupations restricted to $n_i=0,1,2$ and a uniform interaction $U$,
the following operator identities hold:
\begin{equation}
n_i(n_i-1)=2P_{2,i},\qquad
V=\frac{U}{2}\sum_i n_i(n_i-1)=UN_{\rm sites}\hat r.
\label{eq:doublon_interaction_identity}
\end{equation}
At global unit filling,
$\sum_i\bar n_i=N_{\rm sites}$ with $\bar n_i=\langle n_i\rangle$,
\begin{align}\label{eq:doublon_to_hopping}
\frac{1}{N_{\rm sites}}\sum_i\langle(n_i-1)^2\rangle&=2r,\nonumber\\
\frac{1}{N_{\rm sites}}\sum_i\operatorname{Var}(n_i)
&=2r-\frac{1}{N_{\rm sites}}\sum_i(\bar n_i-1)^2.
\end{align}
Thus the site-averaged variance equals $2r$ only when every local mean
density equals one. Open boundaries, disorder, or a density wave can
produce the indicated correction even at fixed total particle number.

For $\hat H=K+V$, where
\begin{equation}
K=-J\sum_{\langle i,j\rangle}(a_i^\dagger a_j+\mathrm{h.c.}),
\end{equation}
any two orthogonal energy eigenstates $|g\rangle$ and $|h\rangle$ obey
$\langle h|\hat H|g\rangle=0$. Hence
\begin{equation}
\langle h|K|g\rangle=-\langle h|V|g\rangle
=-UN_{\rm sites}\langle h|\hat r|g\rangle.
\label{eq:hopping_doublon_matrix_element}
\end{equation}
At nonzero $J$, a uniform fractional hopping modulation gives
$\delta H(t)=[\delta J(t)/J]K$. The same transition matrix element
therefore connects the hopping drive to the doublon readout. Equation~\eqref{eq:hopping_doublon_matrix_element}
holds for any such eigenstate pair; it does not, by itself, identify a
mode as Higgs-like. A uniform potential contributes only a constant in
a fixed-number sector, whereas a nonuniform potential adds its own
off-diagonal matrix element and must be included separately.

For a coherent superposition of two eigenstates in the same number sector,
\begin{equation}
|\psi(t)\rangle=c_g e^{-iE_gt/\hbar}|g\rangle+
c_h e^{-iE_ht/\hbar}|h\rangle,\qquad |c_g|^2+|c_h|^2=1,
\end{equation}
the measured doublon fraction is
\begin{equation}
r(t)=|c_g|^2r_{gg}+|c_h|^2r_{hh}
+2\operatorname{Re}\!\left[c_g^*c_h r_{gh}e^{-i(E_h-E_g)t/\hbar}\right],
\label{eq:doublon_coherent_readout}
\end{equation}
where $r_{\alpha\beta}=\langle\alpha|\hat r|\beta\rangle$ and
$r_{gh}=-\langle g|K|h\rangle/(UN_{\rm sites})$ under the assumptions
above. The oscillation comes from coherence between states with the
same total particle number and reveals their energy difference. An
incoherent mixture has no interference term. A static or slowly drifting
doublon population is therefore not sufficient evidence of a coherently
excited Higgs mode. Its identification also uses the drive symmetry,
spatial response, and frequency, as discussed in Appendix~\ref{app:mf_doublon}
and tested experimentally in Appendix~\ref{app:selectivity}.

Energy absorption under hopping modulation and time-resolved doublon
counting can address the same transitions, but their measured amplitudes
and response functions differ. A cold-atom experiment used tunneling
modulation with a temperature readout~\cite{Endres2012}; trap
inhomogeneity complicates comparison with a uniform system. In a finite
array, the discrete spectrum can restrict decay and resolve separate
resonances, but it does not by itself exclude hybridization, dephasing,
or environmental damping (Appendix~\ref{app:mesoscopic_to_thermodynamic}).

\section{Matrix Product States methods}
\label{sec:mps}

Here we describe our main numerical technique.  We model our quantum states using a Matrix Product State (MPS) ansatz. We sequentially label the sites $1\cdots N$, and define
\begin{align}\label{eq:mps}
|\psi\rangle &= \sum_{s_1\cdot s_N} \sum_{\mu_1\cdots \mu_{n-1}}
A^{[1] s_1}_{\mu_1} A^{[2] s_2}_{\mu_1\mu_2}\cdots A^{[N] s_2}_{\mu_{N-1}} |s_1 s_2\cdots s_N\rangle.
\end{align}
Here, the site indices $s_j$ label states that make up the physical local Hilbert space. In our case $s_j=0,1,2$ represents the number of particles on a site. The bond indices $\mu_j$ can take up to $\chi$ distinct values. This ansatz can describe an arbitrary quantum state if $\chi$ is made sufficiently large. We use standard Density Matrix Renormalization Group (DMRG) algorithms to optimize the coefficients $A^{[j] s_j}_{\mu_{j-1}\mu_j}$ \cite{PhysRevLett.69.2863}.  All calculations make use of the ITensor library \cite{SciPostPhysCodeb.4,SciPostPhysCodeb.4-r0.3}.

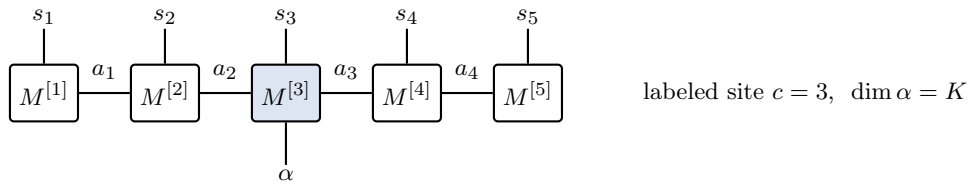
\begin{figure}[H]
\centering
\begin{tikzpicture}[x=16mm,y=10mm]
  \foreach \i in {1,2,4,5} {\node[tens] (A\i) at (\i,0) {$M^{[\i]}$};}
  \node[lbld] (A3) at (3,0) {$M^{[3]}$};
  \foreach \i [evaluate=\i as \j using int(\i+1)] in {1,2,3,4} {\draw[leg] (A\i) -- (A\j);}
  \foreach \i in {1,...,5} {\draw[leg] (A\i) -- ++(0,0.85) node[lab,above] {$s_\i$};}
  \draw[leg] (A3) -- ++(0,-0.95) node[lab,below] {$\alpha$};
  \node[lab] at (1.5,0.28) {$a_1$};
  \node[lab] at (2.5,0.28) {$a_2$};
  \node[lab] at (3.5,0.28) {$a_3$};
  \node[lab] at (4.5,0.28) {$a_4$};
  \node[lab,right=10mm of A5] {labeled site $c=3$, \ $\dim\alpha=K$};
\end{tikzpicture}
\caption{A schematic 5-site chain with orthogonality center at $c=3$. We can choose a state by selecting $ \alpha $, for every $\alpha=1,\dots,K$.  In the mixed canonical form, all tensors to the left of the orthogonality center are left isometries, and those to the right are right-isometries.}\label{exampleMPS}
\end{figure}

To calculate excitations we use a global Krylov method.  To minimize the impact of truncation errors we use a multi-targeted MPS ansatz, where all excitations are described by using a common set of environment tensors \cite{qz9v-p4j5,Baker2024}.  In particular we take one site $c$, and affix a label to it -- taking $A^{[c] }\to A^{[c] \alpha}$ -- where $\alpha$ labels the excitation.  All other tensors are independent of $\alpha$.  In practice the labeled site is the orthogonality center, and we shift the label's location whenever we perform singular value decompositions that move the orthogonality center. Fig.~\ref{exampleMPS} schematically illustrates this structure.

Our goal is to calculate the spectral function
\begin{equation}
  S(\omega) \;=\; \sum_x \bigl| \melem{x}{\Ot}{\psi_0} \bigr|^2\,
                  \delta\bigl(\omega - (E_x - E_0)\bigr),
  \qquad
  \Ot \;\equiv\; O - \melem{\psi_0}{O}{\psi_0},
  \label{eq:spectral}
\end{equation}
where $|\psi_0\rangle$ is the ground state with energy $E_0$.  The probe operator $O$ is either the hopping or potential, depending on if we are modeling the Higgs or Goldstone excitation experiments.  The sum in Eq.~(\ref{eq:spectral}) is over all excited state wavefunction $|x\rangle$, with energy $E_x$.  We approximate this spectral function by generating the Krylov space $\mathcal{K}_m$ spanned by $|v_1\rangle\propto \tilde O |\psi_0\rangle,$ $|v_2\rangle\propto H|v_1\rangle \cdots |v_m\rangle\propto H^{m-1}|v_1\rangle$.  We replace $|x\rangle$ and $E_x$ in Eq.~(\ref{eq:spectral}) by the eigenvalues and eigenvectors of the Hamiltonian projected into this space.  While this construction introduces some approximation, the spectrum found from the exact Krylov space correctly reproduces all moments  $I_n=\int S(\omega) \omega^n d\omega$ for $n<m$.

We use our multitargeted MPS representation to approximate the states in $\mathcal{K}_m$.  Since they share a common environment, linear algebra operations, such as orthogonalization, is efficient on them.  We use a variant of the Lanczos algorithm to generate the Krylov space.  Beginning with $|\phi_1\rangle=|v_1\rangle=\tilde O |\psi_0\rangle$, we generate the environment which spans $|v_1\rangle$, and $H |v_1\rangle$.  We use a Gram-Schmidt procedure to create $|\phi_2\rangle$ which is orthogonal to $
|\phi_1\rangle$.  We repeat, at the $j$'th step applying $H$ to $|\phi_j\rangle$, expanding the environment, then orthogonalizing with previous wavefunctions. The resulting Hamiltonian, projected into the Krylov space, is tri-diagonal.  

An important challenge is that our MPS representation results in approximations.
Crucially, the orthogonality between the states in our basis is lost when we project into the new environment.  We mitigate these errors by reorthogonalizing all states after each step.  A second consequence of these approximations is that generically $|\phi_j\rangle$ will have some weight which lies outside of the exact Krylov space.  These components can grow as $H$ continues to act on the state, and after many iterations the resulting variational manifold may have very little overlap with the exact Krylov space.  We use a deflation technique to remove these unwanted components.  After $m$ steps we construct the projected Hamiltonian, and find its eigenvalues $\lambda_j$ and eigenvectors $|\chi_j\rangle$ (typically referred to as the Ritz value and Ritz vectors).  For each eigenvector we calculate 
\begin{align}
c_j &= \langle \chi_j|\phi_1 \rangle
=\langle \chi_j |\tilde O |\psi_0\rangle
\end{align}
If $c_j=0$, then the eigenvector is orthogonal to the Krylov space, and does not contribute to the spectral density in Eq.~(\ref{eq:spectral}).  If $|c_j|^2$ is smaller than a threshold, we use a shifted QR to remove the state from our space.  Without this deflation step we would be restricted to small $m$.

An eigenvector, $|\chi_j\rangle$,  of the Hamiltonian projected into $\mathcal{K}_m$, has converged when $H|\chi_j\rangle=\lambda_j |\chi_j\rangle$, where $H$ is the unprojected Hamiltonian.  This condition would be true if $\langle \phi_m|\chi_j\rangle=0$.  Thus a convenient metric for convergence is the total spectral weight outside of the Krylov space,
\begin{align}
R &= \sum_j |\langle \phi_m|\chi_j\rangle|^2 |\langle \phi_1|\chi_j\rangle|^2
\end{align}
We iterate until $R$ falls below a threshold.

\section{Thermodynamic
limit}\label{app:mesoscopic_to_thermodynamic}

\begin{figure}
    \centering
    \includegraphics[width=1.0\linewidth]{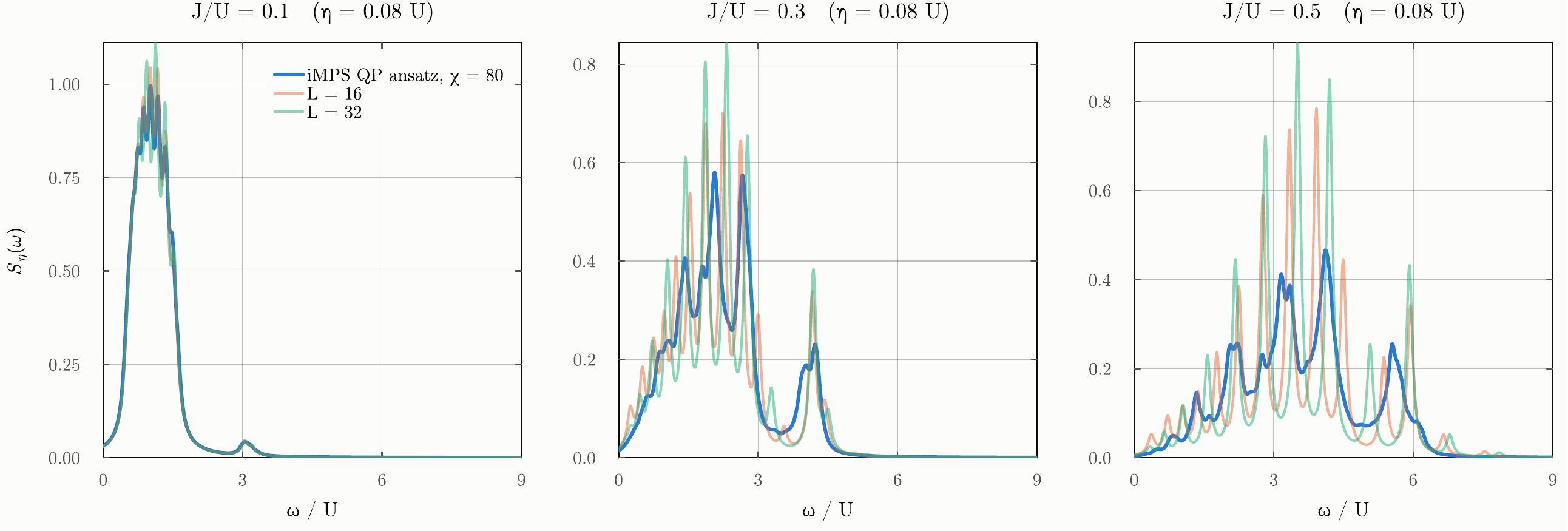}
    \caption{Kinetic spectral function $S_\eta(\omega)$, Eq.~\eqref{eq:broadened_spectral}, at fixed small broadening $\eta = 0.08\,U$ for $J/U = 0.1$, $0.3$ and $0.5$ (left to right). Translucent
    curves are finite chains obtained by Krylov expansion at $\chi = 2000$, $L = 16$ (orange) and $L = 32$ (green); the solid blue curve is the iMPS quasiparticle ansatz at $\chi = 80$. At $J/U = 0.1$ the weight is concentrated in a single narrow feature near $\omega \simeq U$ and the three
    calculations already coincide. As $J/U$ increases the weight spreads over a wider window and the node density grows, so at fixed $\eta$ the individual nodes remain resolved and the finite-size spectra oscillate strongly about the smooth infinite-system lineshape --- an artefact of level discreteness rather than a physical difference.}
    \label{fig:Spectral_Function_Nodes}
\end{figure}

\begin{figure}
    \centering
    \includegraphics[width=1.0\linewidth]{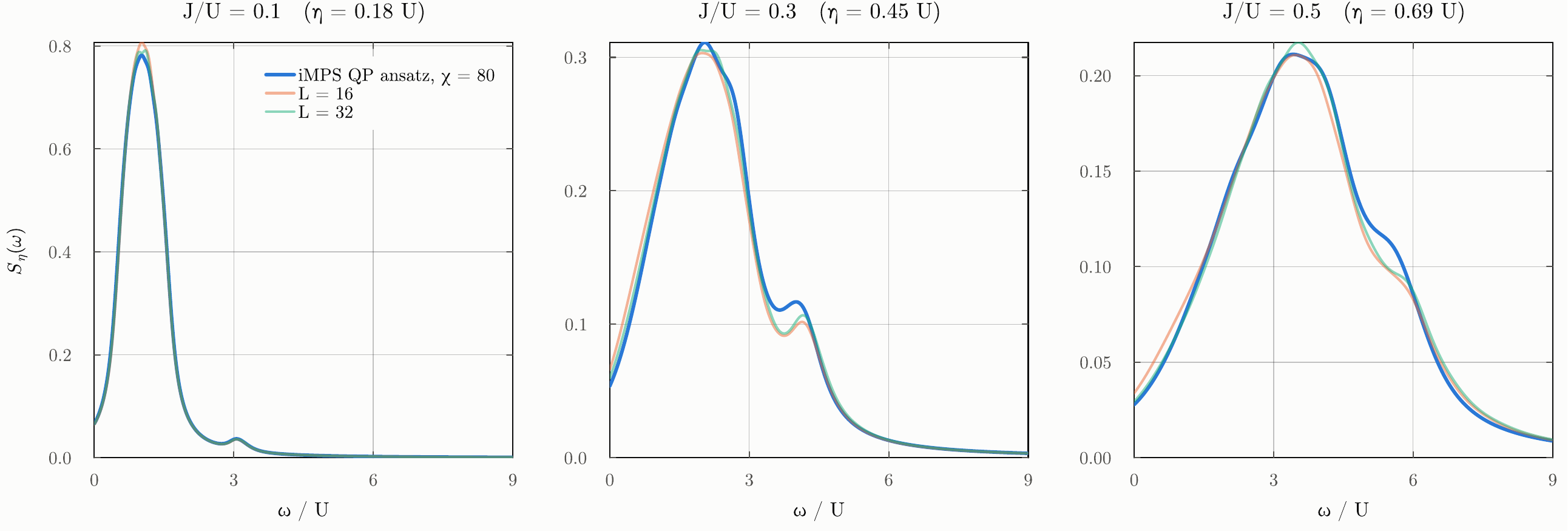}
    \caption{Same data as Fig.~\ref{fig:Spectral_Function_Nodes}, with $\eta$ set to the mean level spacing at each coupling ($\eta = 0.18$, $0.45$ and $0.69\,U$ for $J/U = 0.1$, $0.3$, $0.5$). 
    }
    \label{fig:Spectral_Function_Broadened}
\end{figure}

Here we present results from our numerical modeling which illustrates the connection between the sharp Higgs modes observed in our mesoscopic system and the broad features expected in the thermodynamic limit.  For these calculations we consider a one dimensional geometry, where we have efficient numerical techniques for studying large systems. We emphasize that the experiment is two-dimensional, and the detailed structure of the modes has significant dependence on dimension.  Nonetheless we can use these one-dimensional calculations to illustrate the key features, principally arising from the fact that the Higgs mode is coupled to the multi-Goldstone excitations.  In a small system this means that when one modulates the hopping strength one finds a series of sharp resonances.  As one increases the system size these individual peaks coalesce into a broad spectral feature.  The principle role of dimensionality is to control the shape of that feature.

We use two numerical approaches, both based upon matrix product state ansatzes.  For our finite size calculations we use the techniques described in Appendix~\ref{sec:mps}.  We work with a one dimensional chain of length $L=16,32$, and calculate the spectral function in Eq.~(\ref{eq:spectral}) with $O=K$ corresponding to modulating the hopping.  This excites the Higgs mode.  We also perform calculations in the thermodynamic limit, using the VUMPS algorithm~\cite{PhysRevB.97.045145} to optimize a translationally invariant infinite matrix product state ansatz, which is the generalization of Eq. (\ref{eq:mps}) to an infinite number of sites:  Each of the $A$ matrices are identical, and chosen to minimize $\langle \psi|H |\psi\rangle$.  We calculate the excitations by using the $k=0$ quasiparticle ansatz ~\cite{PhysRevB.85.100408}, corresponding to an equal weighted sum of wavefunctions where a single $A$ is replaced by a different matrix $B$.  We find the eigenvalues of the Hamiltonian projected into this variational space, and use them to approximate the response function, $S(\omega)$.  The variational space is $\chi$-dimensional, corresponding to the bond-dimension of the matrix product state, and the spectrum therefore consists of $\chi$ discrete peaks.  As $\chi\to\infty$ it converges to a continuum.

Figure~\ref{fig:Spectral_Function_Nodes} shows the resulting spectral functions, with the delta-functions broadened into Lorentzians of with $\eta=0.08 U$, 
\begin{equation} \label{eq:broadened_spectral}
    S_\eta(\omega) = \frac{\eta}{\pi}\sum_x
    \frac{\bigl|\langle\psi_0|\tilde{K}\vert{}x\rangle\bigr|^{2}}
         {(\omega-E_x+E_0)^{2}+\eta^{2}}
\end{equation}
Here $\tilde K=K-\langle \psi_0|K|\psi_0\rangle$.
The three panels correspond to the Mott regime, $J/U=0.1$, near criticality, $J/U=0.3$, and deep in the superfluid regime $J/U=0.5$.  In all cases the infinite MPS data is shown, as well as the finite size results.  The same data is shown in Fig.~\ref{eq:broadened_spectral} with larger broadening, corresponding to the mean level spacing.
Importantly, the broadened spectra all agree with one-another.

There are a number of noteworthy features of Fig.~\ref{fig:Spectral_Function_Nodes}.  In the Mott regime the spectrum is dominated by a peak at near $\omega=U$, corresponding to the energy for creating  a doublon.  There is a much smaller feature near $\omega=3 U$, corresponding to forming triplons.  While the spectra become more complicated at larger $J/U$, these two features continue to dominate the envelope.  At larger $J/U$ the individual peaks spread out.  In the finite $L$ case the discreteness is due to finite size effects:  The excitations have a component corresponding to pairs  of phonons with momentum and energy.  In the infinite system the discreteness is an artifact of the quasiparticle ansatz with finite bond dimension.  The ansatz has a finite correlation length, which sets a maximum separation for the phonon pairs.  This leads artificial constraint leads to a non-physical quantization of the allowed energies.

\section{Finite size scaling}

\begin{figure}[h]
    \centering
    \includegraphics[width=\textwidth]{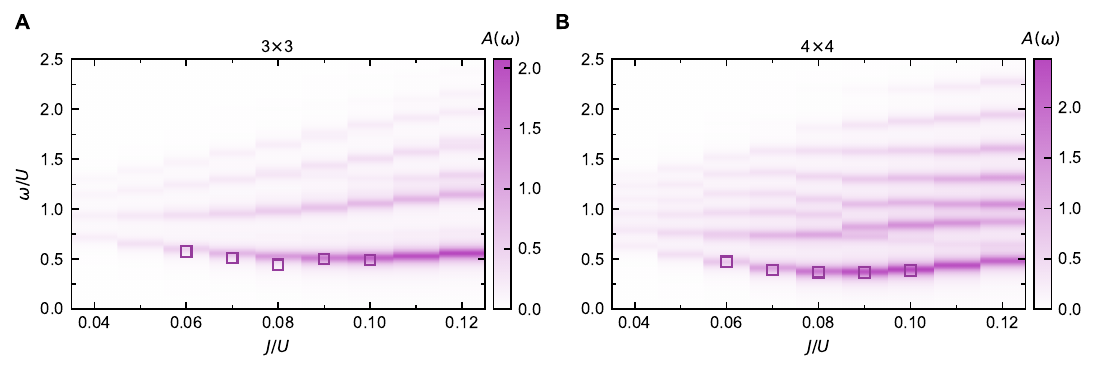}
    \caption{
    \textbf{Simulated kinetic spectral function for two system sizes.} Unit normalized broadened spectral function ($\eta=0.05$). The $4\times 4$ plot is reproduced from the main text Fig.~\ref{fig:fig_higgs_spectrum_theory}. Experimental peak locations from vacuum Rabi spectroscopy are shown as square markers.
    }\label{fig:fig_theory_higgs_spectrum}
\end{figure}

Here we report on our numerical calculations of the kinetic spectral density, Eq.~(\ref{eq:broadened_spectral}), corresponding to modulating the hopping, for $3\times 3$ and $4\times 4$ lattices, and a range of $J/U$.  The results are summarized in Fig.~\ref{fig:fig_theory_higgs_spectrum}.  Similar to the 1D calculations of Appendix~\ref{app:mesoscopic_to_thermodynamic}, we see a sequence of discrete peaks, whose spacing falls as the system size increases.  In the 2D case however, the spectral weight is more concentrated to lower energies than in 1D, with the first peak being distinctly stronger than the others.  The lowest energy mode has a distinct minimum, which marks the crossover between Mott insulating and superfluid regimes.

The excited peaks in Fig.~\ref{fig:fig_theory_higgs_spectrum}
occasionally show dislocations, where the frequencies jump when $J/U$ is changed.  These are likely artifacts, caused by the truncation errors resulting from expressing the quantum states as matrix product states with finite bond dimension.  Despite these features, the envelope and rough mode spacing should be robust.

\putbib[bibliography] %
\end{bibunit}
\end{appendices}

\end{document}